\documentclass[12pt,preprint]{aastex}
\usepackage{lscape}
\usepackage{multirow}
\usepackage{booktabs}
\usepackage{graphicx}
\usepackage{amsmath}
\usepackage{longtable}
\usepackage{rotating}
\usepackage{enumerate}

\newcommand{\widthsigma}{w_\Sigma}
\newcommand{\widthsigmanorm}{\Delta_\Sigma}
\newcommand{\depthsigma}{\delta_\Sigma}
\newcommand{\widthimage}{w_I}
\newcommand{\widthimagenorm}{\Delta_I}
\newcommand{\depthimage}{\delta_I}
\newcommand{\msun}{M_\odot}
\newcommand{\mj}{M_{\rm J}}
\newcommand{\ms}{M_{\rm S}}
\newcommand{\mn}{M_{\rm N}}
\newcommand{\mdisk}{M_{\rm disk}}
\newcommand{\rsun}{R_\odot}
\newcommand{\rsub}{r_{\rm sub}}
\newcommand{\rp}{r_{\rm p}}
\newcommand{\rmin}{r_{\rm min}}
\newcommand{\rminsigma}{r_{\rm min,\Sigma}}
\newcommand{\rminimage}{r_{\rm min,I}}

\newcommand{\mplanet}{M_{\rm p}}

\begin{document}
\title{What is the Mass of a Gap-Opening Planet?}

\shorttitle{Gaps in NIR Images}

\shortauthors{Dong \& Fung}

\author{Ruobing Dong\altaffilmark{1} \& Jeffrey Fung\altaffilmark{2,3}}
\altaffiltext{1}{Steward Observatory, University of Arizona, rdong@email.arizona.edu}
\altaffiltext{2}{Department of Astronomy, University of California, Berkeley}
\altaffiltext{3}{NASA Sagan Fellow}

\clearpage

\begin{abstract}

High contrast imaging instruments such as GPI and SPHERE are discovering gap structures in protoplanetary disks at an ever faster pace. Some of these gaps may be opened by planets forming in the disks. In order to constrain planet formation models using disk observations, it is crucial to find a robust way to quantitatively back out the properties of the gap-opening planets, in particular their masses, from the observed gap properties, such as their depths and widths. Combing 2D and 3D hydrodynamics simulations with 3D radiative transfer simulations, we investigate the morphology of planet-opened gaps in near-infrared scattered light images. Quantitatively, we obtain correlations that directly link intrinsic gap depths and widths in the gas surface density to observed depths and widths in images of disks at modest inclinations under finite angular resolution. Subsequently, the properties of the surface density gaps enable us to derive the disk scale height at the location of the gap $h$, and to constrain the quantity $\mplanet^2/\alpha$, where $\mplanet$ is the mass of the gap-opening planet and $\alpha$ characterizes the viscosity in the gap. As examples, we examine the gaps recently imaged by VLT/SPHERE, Gemini/GPI, and Subaru/HiCIAO in HD 97048, TW Hya, HD 169142, LkCa~15, and RX J1615.3-3255. Scale heights of the disks and possible masses of the gap-opening planets are derived assuming each gap is opened by a single planet. Assuming $\alpha=10^{-3}$, the derived planet mass in all cases are roughly between 0.1--1 $\mj$.

\end{abstract}



\section{Introduction}\label{sec:intro}

In the past few years, high angular resolution observations of protoplanetary disks have revolutionized our understanding of planet formation. Rich structures have been identified in disk images in near infrared (NIR) and in mm/cm dust continuum and gas emission. Among them, a particularly intriguing class of features is narrow gaps, found in systems such as TW Hya \citep{debes13, rapson15twhya, akiyama15, andrews16}, HD 169142 \citep{quanz13gap, momose15}, HD 97048 \citep{ginski16, walsh16, vanderplas16}, AB Aur \citep{hashimoto11}, V 4046 Sgr \citep{rapson15v4046}, HD 141569 \citep{weinberger99, mouillet01, konishi16, perrot16}, RX J1615.3-3255 \citep{deboer16}, and HL Tau \citep{brogan15, yen16, carrascogonz16}. These gaps have both their inner and outer edges revealed in images, enabling a full view of the gap structure and particularly, a measurement of the gap width. Numerical studies of observational signatures of planet-induced structures have suggested that gaps may be opened by planets \citep[e.g.][]{wolf05, jangcondell12, jangcondell13, fouchet10, gonzalez12, ruge13, pinilla15twoplanets,pinilla15hd100546, dong15gaps, jin16, dong16armviewing}. While giant gaps, found in systems such as transitional disks \citep[e.g.][]{hashimoto12, follette15, stolker16sao206462}, may be common gaps opened by multiple planets \citep{zhu11, dodsonrobinson11, duffell15dong}, the above mentioned narrow gaps are more likely to be the products of single planets.

The masses and locations of planets still forming in disks are key quantities in the study of planet formation. Different formation scenarios, such as ``core accretion'' and ``gravitational instability'' models, predict planets to form at different locations, with different final masses. Measuring the masses of gap-opening planets ($\mplanet$) will thus help us distinguish different planet formation scenarios. To do this, quantitative connections between $\mplanet$ and gap properties that are directly measurable from images must be established. At the moment, a general scheme for this purpose is lacking. \citet{rosotti16} pioneered in this direction by connecting $\mplanet$ with gap width at both infrared and millimeter (mm) wavelengths. They explored a limited parameter space (e.g., no variation on disk viscosity), and it is unclear whether synthetic observations based on 2D hydro simulations employed by \citet{rosotti16} generate the same results as 3D hydro simulations.

In this work, we propose to advance this field by connecting $\mplanet$ with gap properties in NIR scattered light. Our goal is to establish direct and quantitative relations between two observables --- the gap depth ($\depthimage$; the contrast) and width ($\widthimage$; the radial extent) --- with $\mplanet$, and two disk parameters, the aspect ratio $h/r$ and $\alpha$-viscosity \citep{shakura73}. This effort is newly motivated by recent advances in NIR polarized scattered light imaging driven by Subaru/HiCIAO \citep{tamura06}, Gemini/GPI \citep{macintosh08}, VLT/NACO \citep{lenzen03} and SPHERE\citep{beuzit08}. At the moment these high resolution imaging instruments are discovering gaps at an ever faster pace, which urgently necessitates the effort presented here.

This $[\depthimage,\widthimage]-[\mplanet,h/r,\alpha]$ connection can be segmented into two parts,
\begin{enumerate}
\item $[\depthimage,\widthimage]-[\depthsigma,\widthsigma]$: connect gap depth and width in images to the {\it physical} gap depth and width in the gas distribution, $\depthsigma$ and $\widthsigma$, measured from the disk surface density $\Sigma$;
\item $[\depthsigma,\widthsigma]-[\mplanet,h/r,\alpha]$: connect $\depthsigma$ and $\widthsigma$ to $\mplanet$, $h/r$, and $\alpha$. 
\end{enumerate}

The second part has been well studied both analytically and numerically. Analytically, \citet[see also \citealt{fung14, duffell15gap}]{kanagawa15} have shown that based on torque balancing one can derive a simple scaling relation
\begin{equation}
\depthsigma-1=\frac{\Sigma_0}{\Sigma_{\rm min}}-1\propto\frac{q^2(h/r)^5}{\alpha^2},
\label{eq:gapdepthsigma}
\end{equation}
where $\Sigma_{\rm min}$ and $\Sigma_0$ are the depleted and initial (undepleted) surface density in the gap, and $q=\mplanet/M_\star$. For the gap width, \citet{goodman01} calculated the propagation of planet-induced density waves, and showed that in the weakly-nonlinear regime the gap width, the waves dissipate and deposit angular momentum over the length scale ``shocking length'', which is proportional to the disk scale height. It is therefore likely that the width of the gap will also be proportional to the scale height:
\begin{equation}
\widthsigma=r_{\rm out}-r_{\rm in}\propto h,
\label{eq:width-h}
\end{equation}
where $r_{\rm out}$ and $r_{\rm in}$ are the radius of the outer and inner gap edges, respectively. This relation will be tested and confirmed with our models. The weakly-nonlinear regime applies to planets with masses comparable or smaller than the thermal mass $M_{\rm thermal}$,
\begin{equation}
\mplanet\lesssim M_{\rm thermal}=M_\star\left(\frac{h}{r}\right)^3.
\label{eq:mthermal}
\end{equation}

Several groups have carried out numerical simulations to fit $\depthsigma$ and $\widthsigma$ as power laws of $\mplanet$, $h/r$, and $\alpha$ \citep{duffell13, fung14, duffell15gap, kanagawa15, kanagawa16width}. These empirical correlations were synthesized from 2D hydro simulations, and \citet{fung16} confirmed that gap opening in 3D produces the same gap profiles as in 2D. The exact functional forms vary somewhat in the literature, mostly due to variations in the definitions of $\widthsigma$ and $\depthsigma$. For the gap depth, \citet{fung14} found, for deep gaps with $\depthsigma$ exceeding 10, up to $10^4$, Equation~\ref{eq:gapdepthsigma} agrees with simulations to within a factor of 2. For the gap width, hydro simulations by \citet{muto10}, \citet{dong11linear, dong11nonlinear}, \citet{zhu13}, and \citet{duffell13} have verified the results of \citet{goodman01}, and showed that gap opening indeed initially occurs at a fixed number of scale heights away from the planet's orbit for a given planet mass. By defining the gap width as the radial distance between the two edges where the surface density drops to 50\% of the initial value, \citet{kanagawa16width} found $\widthsigma\propto q^{1/2}(h/r)^{-3/4}\alpha^{-1/4}$, which not only has dependencies on $\mplanet$ and $\alpha$, but also predicts an inverse relation with $h/r$. Whether these differences reflect some unknown physics of of gap opening, or simply due to the different definitions of $\widthsigma$, is unclear.

The first step, finding out how a density gap looks like in scattered light images, is the aim of this paper. We address this question by carrying out 2D and 3D hydro and 3D radiative transfer simulations to study synthetic images of planet-opened gaps with parametrized $h/r$ and $\alpha$. This step is affected by properties of the disk and observational conditions, in particular, the angular resolution $\eta$ (a known parameter for a specific observation) and the inclination of the system $i$. The dependence $i$ can be eliminated by only measuring the scattered light radial profile at $\sim90^\circ$ scattering angle (Section~\ref{sec:inclinations}), such as in face-on disks and along the major axis of inclined disks. A few other factors, including the total disk mass $\mdisk$, wavelength $\lambda$, and grain properties, may affect the conversion as well; however, within reasonable ranges in the parameter space assumed in this paper (optically thick, gravitationally stable disks; $0.5\micron\leq\lambda\leq2\micron$; close to $90^\circ$ scattering angle) they are generally unimportant \citep[e.g.,][]{dong12cavity}.  For extremely faint gaps in which the gap bottom reaches the noise level in images, the sensitivity of the observations (i.e., detection limit) will affect the measurement of $\depthimage$; in this work we focus on relatively low mass planets and shallow gaps and do not consider such situation.

We focus on narrow gaps ($\Delta_{\rm gap}<0.5\rp$) opened by a single relatively low mass giant planet about Neptune ($\mn$) and Saturn ($\ms$) mass, up to one Jupiter ($\mj$) mass. The motivation is mainly threefold. First, such planets are probably more common than their more massive siblings \citep{cumming08, brandt14}, and therefore it may be more likely to see gaps opened by such planets. \citet{bowler16} concluded that only $\sim0.6\%$ of  $0.1-3$ $\msun$ stars have $5-13$ $\mj$ planets at orbital distances of $30-300$ AU. Although little is known about the statistics of sub-Jovian planet at tens of AU, observations such as HL Tau hints at their existence \citep{jin16}. Second, massive planets ($\gtrsim\mj$) may open eccentric gaps \citep[e.g.,][]{kley06, dangelo06}, and vortices may form at the gap edge excited by the Rossby wave instability \citep[e.g.,][]{li00, lin10, zhu14votices}. These features complicate the interpretation of gaps in observations. Third, wide gaps may be common gaps opened by multiple planets, in which case there is a degeneracy between the number of planets and their masses \citep{dong16td}, preventing unique solutions on planets masses. The lower limit of $\mplanet$ in our work is set so that its gap may be visible in scattered light \citep[e.g.,][]{rosotti16}.


\section{Simulation Setup}\label{sec:setup}

In this section, we briefly introduce the numerical methods employed to produce synthetic images of gaps. In short, we use the hydrodynamics code \texttt{PEnGUIn} \citep{fung15thesis} to calculate the density structures of planet-opened gaps in 2D and 3D, then ``translate'' the resulting density structures to scattered light images using radiative transfer simulations by the Monte Carlo \citet{whitney13} code. For hydro simulations we follow the procedures described in \citet{fung14}, while for radiative transfer simulations we adopt the methods in \citet{dong15gaps, dong16armviewing}. Below we briefly summarize the key processes.

\subsection{Hydrodynamical Simulations}\label{sec:hydro}

The 2D simulations performed in this paper have an identical setup as the one used in \citet{fung14}, and the 3D one is identical to \citet{fung16}. The only differences are that we choose the parameters $q$, $h/r$, and $\alpha$ from a different parameter space, focusing more on smaller planets and lower disk viscosities, and that all results here are obtained after 6000 orbits, equivalent to about 1 Myr at 30 AU around a solar-mass star. This time is sufficient for models with $\alpha\geq 10^{-3}$ to reach a quasi-steady state. For the models with the lowest $\alpha$ values, the gaps have not yet fully settled, but they are still relevant models since 1 Myr is already a significant fraction of the typical protoplanetary disk lifetime. We recapitulate some important features here, and refer the reader to \citet{fung14} and \citet{fung16} for the details.

The initial disk profile assumes the following surface density and sound speed profiles:
\begin{equation}\label{eqn:initial_sigma}
\Sigma_0 =  \Sigma_{\rm p} \left(\frac{r}{r_{\rm p}} \right)^{-\frac{1}{2}}\, ,
\end{equation}
\begin{equation}\label{eqn:initial_c}
c_{\rm s} =  c_{\rm p} \left(\frac{r}{\rp} \right)^{-\frac{1}{2}}\, ,
\end{equation}
where we set $\Sigma_{\rm p} = 1$,\footnote{Since we do not consider the self-gravity of the disk, this normalization has no impact on our results.}, and $c_{\rm p}$ is a parameter we vary to obtain different $h/r$ values. Since the disk scale height $h$ is $c_{\rm s}/\Omega_{\rm k}$ with $\Omega_{\rm k}$ being the Keperlian frequency, $h/r$ is constant in radius. In 3D hydrostatic equilibrium, the initial density structure is:
\begin{equation}\label{eqn:initial_rho}
\rho_0 = \rho_{\rm p} \left(\frac{r}{\rp}\right)^{-\frac{3}{2}} \exp\left(\frac{GM}{c_{\rm s}^2} \left[\frac{1}{\sqrt{r^2 +z^2}}-\frac{1}{r}\right]\right) \,,
\end{equation}
where $\rho_{\rm p} = \Sigma_{\rm p}/\sqrt{2\pi h_{\rm p}^2}$, with $h_{\rm p}$ being the scale height at $r=\rp$, is the initial gas density at the location of the planet. Density and sound speed are related by a locally isothermal equation of state. The kinematic viscosity in the disk is $\nu = \alpha c_{\rm s} h$, where $\alpha$ is constant in radius. The simulation grid spans $0.4~\rp$ to $2~\rp$ in radius, which corresponds to 12 AU to 60 AU, if the planet is placed at 30 AU.

Table~\ref{tab:models} lists the parameters and results for all models.

\subsection{Monte Carlo Radiative Transfer Simulations}\label{sec:mcrt}

For radiative transfer calculations, we assume a central source of 2.325 $\rsun$ and 4350 K, appropriate for a 1 $\msun$ star at 1 Myr \citep{baraffe98}. For 3D hydro simulations, we directly feed the density grid into radiative transfer simulations; for 2D hydro simulations, we puff up the disk surface density in the vertical direction $z$ by a Gaussian profile,
\begin{equation}
\rho(z)=\frac{\Sigma}{h\sqrt{2\pi}} e^{-z^2/2h^2}, \label{eq:gauss}
\end{equation}
using the same scale height $h$ as in the hydro simulations. The MCRT simulation domain spans $\pm30^\circ$ from the disk midplane, which is about a factor of 2 higher than the scattered light surface. A 1 AU radius circumplanetary region centered on the planet is excised (except in Section~\ref{sec:2d3d} and Figure~\ref{fig:2d3d}, see below), for reasons laid out in Section~\ref{sec:2d3d}.

The hydro simulations are gas only, while scattered light comes from the dust. We convert gas into dust by assuming the two are well mixed, and assume interstellar medium grains \citep{kim94} for our dust model. These grains are sub-$\micron$ in size, and their scattering properties are calculated using Mie theory. They are small enough that their stopping time is very short compared to their orbital timescale and vertical stirring timescale (characterized by the viscous parameter $\alpha$) in our simulations, so they are dynamically well coupled with gas \citep[e.g.,][]{zhu12}.

We scale the size of the hydro grid so that the planet is at 30 AU. Given the inner working angles (IWA) in current GPI and SPHERE observations ($\sim0\arcsec.1=14$AU at 140 pc), 30 AU is about as close as a planet can be for the inner gap edge to still be outside the IWA. The inner dust disk edge is taken to be dust sublimation radius $\rsub$, where dust reaches temperature of 1600 K; between $\rsub$ and the inner boundary of hydro simulations (12 AU), we assume an axisymmetric inner disk smoothly joining the outer hydro disk with a constant surface density. The details of the inner disk does not matter as long as it is axisymmetric and does not cast significant shadows on the other disk. We scale the hydro disk so that the initial condition $\Sigma_0(r)=10 (r/30{\rm AU})^{-1/2}$ g cm$^{-2}$, resulting in a final total dust disk mass between $5\times10^{-5}$ to $10^{-4}\msun$ inside 60 AU (depending on how deep the gap is). We note that gap width and contrast do not depend on the assumed disk mass within one order of magnitude from the chosen value. Radiative transfer simulations produce full resolution polarized intensity (PI) images at $H$-band, which are convolved by a Gaussian point spread function to achieve a desired angular resolutions. Fiducially we choose $\eta=0\arcsec.04$, comparable to what is achievable by SPHERE and GPI,  unless otherwise noted.

\begin{figure}
\begin{center}
\includegraphics[trim=0 0 0 0, clip,width=0.35\textwidth,angle=0]{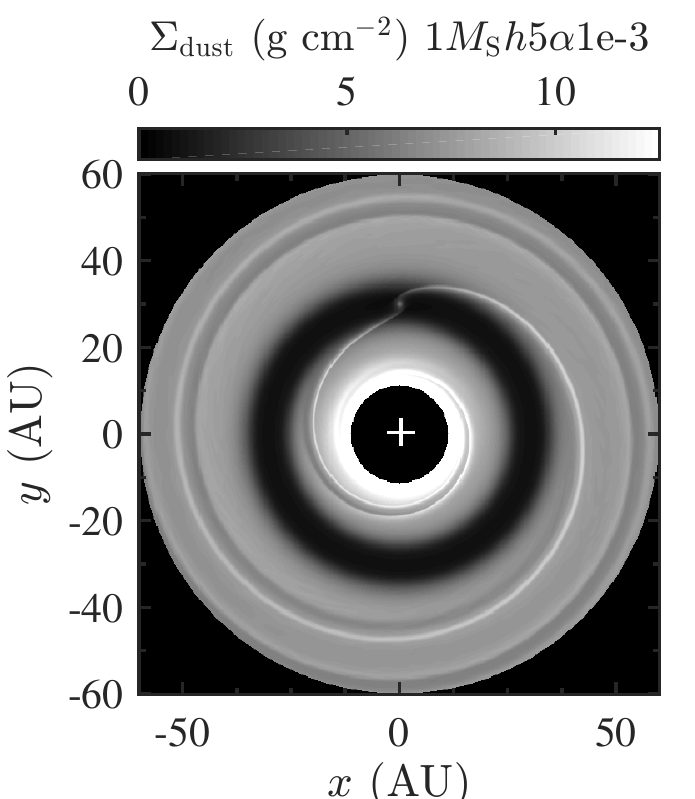}
\includegraphics[trim=0 0 0 0, clip,width=0.9\textwidth,angle=0]{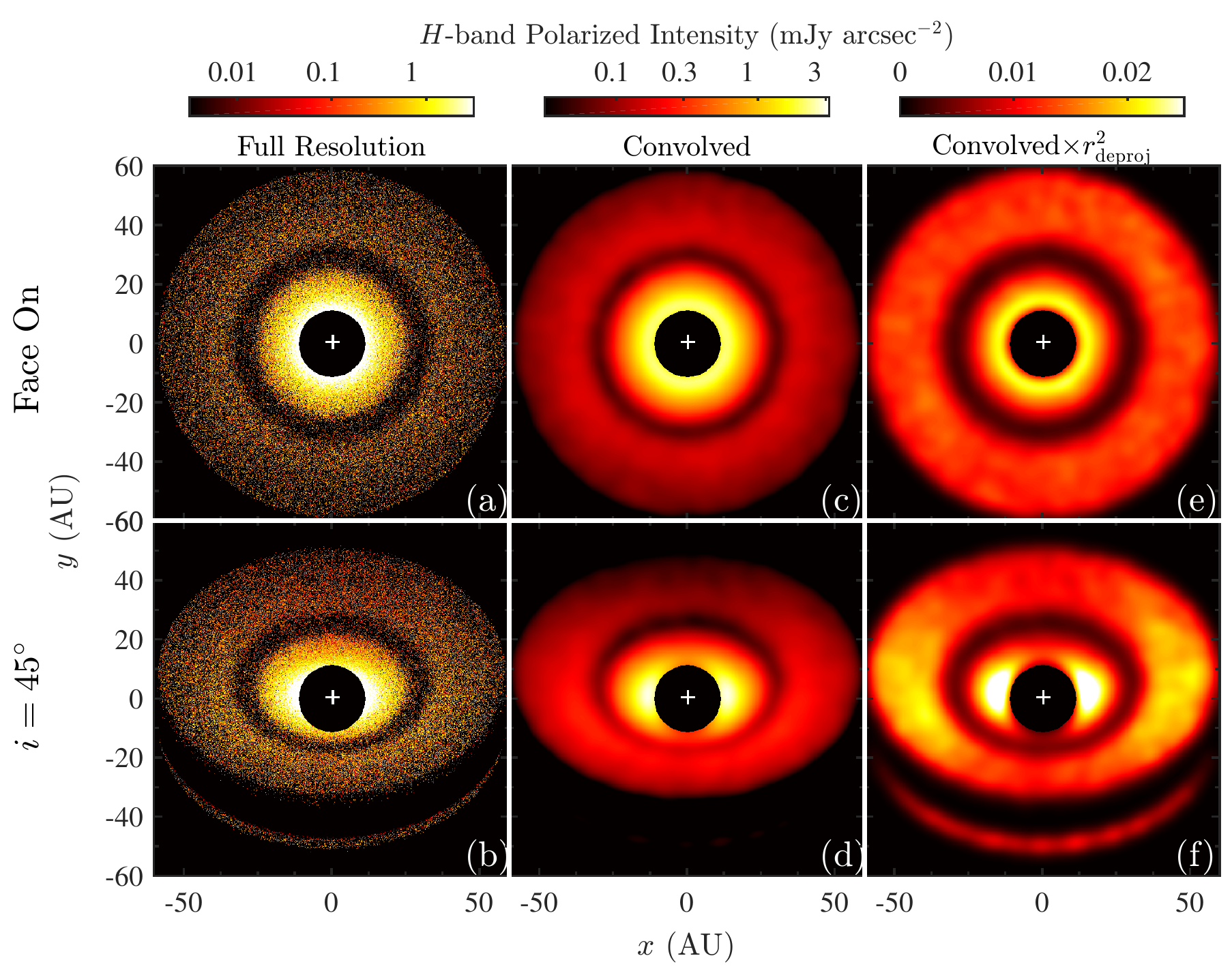}
\end{center}
\figcaption{Dust surface density (top) and $H$-band polarized intensity images at face-on (a,c,d) and $i=45^\circ$ inclinations (b,d,f) of the 1$\ms h$5$\alpha$1e-3 model. The image panels are full resolution images (a,b), convolved images with $\eta=0\arcsec.04$ (c,d), and convolved images scaled by $r^2$ (e,f; $r$ is deprojected distance from the star). See Section~\ref{sec:mcrt} for details.
\label{fig:example}}
\end{figure}

As examples, Figure~\ref{fig:example} shows the dust surface density and scattered light images at both face on and $i=45^\circ$ inclination for Model 1$\ms h$5$\alpha$1e-3. Unless otherwise noted, all images in this paper show $H$-band polarized intensity $I$, although we note that our results generally apply to total intensity images as well (see below). In panel (e) and (f), we scale the images by $r^2$ to compensate for the distance from the star, where $r$ is the distance from the star (deprojected in inclined cases). Unless otherwise noted, all $H$-band images and their radial profiles below have been scaled by $r^2$.

\subsection{2D and 3D Gaps are the same in Scattered Light}\label{sec:2d3d}

\citet{fung16} have shown that the surface density of planet-opened gaps in 3D hydro simulations are nearly the same as gaps in 2D simulations. In this section we show that synthetic images are also practically identical between a 3D model and a model with 2D surface density manually puffed up using Equation~\ref{eq:gauss}. 

Figure~\ref{fig:2d3d} shows two face-on synthetic images and their azimuthally averaged radial profile for the model 1$\ms h$5$\alpha$1e-3. Panel (a) is from a 3D hydro simulation, while panel (b) is produced by puffing up the surface density of the model by the same assumed $h/r$ profile as in the 3D hydro calculation. Note that unlike all other images in the paper, in both panels the circumplanetary region is not excised. The gap profiles (c) from the two images are essentially identical. This result means vertical kinematic support is unimportant in the vertical distribution of material in the gap region, and we can safely puff up 2D hydro surface density structures in MCRT simulations to study gaps in scattered light.

The differences between the two images are in the local structures. First, the spiral arms are clearly more prominent in (a) than in (b), due to vertical kinematic support in planet-induced density waves \citep[Dong \& Fung, submitted]{zhu15densitywaves}. Second, the circumplanetary region sticks out in (b) while it is not noticeable in (a). This is because when puffing up 2D structures into 3D, the circumplanetary region, which has a higher surface density than the surroundings, is artificially expanded in the vertical direction by the same $h/r$ as the rest of the disk. In reality though the gravity in the circumplanetary region is dominated by the planet so the actual $h/r$ of the circumplanetary region is much smaller. This leads to artificially brightening of the region in scattered light in (b). For this reason, and to avoid the artificial shadow in the outer disk produced by this structure \citep[e.g.,][]{jangcondell09}, we excise the circumplanetary region in 2D surface density maps when puffing it up in MCRT simulations in the rest of the paper.


\section{The Conversions}\label{sec:results}

In this section, we explore the connections between gap properties in surface density to disk and planet parameters $h/r$, $\alpha$, and $\mplanet$, and the correlations between gaps in scattered light and gaps in surface density. 

\subsection{The Qualitative Picture}\label{sec:qualitative}

Figure~\ref{fig:sigma_image_rp} shows the radial profiles of the surface density, and both full resolution and convolved images at face-on viewing angle for three representative models with $\alpha=10^{-3}$, $h/r=0.05$, and $\mplanet=2\mn$, $1\ms$, and $1\mj$. The images have been scaled by $r^2$. Qualitatively, in all three cases, each with very different gap depths, the radial profiles of the full resolution images closely traces the variations in the surface density profiles. This general behavior has been found by \citet{muto11} using analytical calculations of scattered light grazing angles in disks with narrow gaps. The two deviate near the gap bottom, where the scattered light profile does not reach as deep as the surface density. Also, the outer edge of the scattered light gap lies slightly inside the surface density gap. 

These observations can be qualitatively explained. Approximately, scattered light comes from a surface with optical depth of unity to the star \citep{takami14}. For disks with constant $h/r$, if the surface density variation is smooth, this surface is flat at roughly a constant polar angle $\theta_{\rm surface}\propto h/r$ (i.e., not concave-up as in flared disks). In this case, at $i=0$ the $r^2$-scaled intensity of the scattered light at each radius is roughly proportional to the surface density of the dust above the surface, $I\propto\Sigma_{\theta>\theta_{\rm surface}}(r)$. Because $\Sigma_{\theta>\theta_{\rm surface}}(r)\propto\Sigma(r)$ for disks with constant $h/r$, it follows that $I\propto\Sigma(r)$, which is what we find. This relation breaks when the surface density fluctuation is so large that the disk surface is no longer at roughly a constant $\theta_{\rm surface}$, which happens when the gap depletion is high, and at the gap edges. Also, this analysis does not formally hold if the disk surface is flared (i.e., $h/r$ increases outward). However, gaps are local structures, and the global flareness of the disk may not significantly affect the width and depth of local structures. We will come back to this point later in Section~\ref{sec:flareness}.

Convolving the full resolution image by a finite PSF smooths structures. Convolved images (dotted line) thus closely follow full resolution images outside the gap, as the variations occur on spatial scale much larger than the angular resolution ($\sim6$AU for $\eta=0\arcsec.04$ and $d=$140 pc). Inside the gaps, convolved images have less steep edges and shallower depths.

\subsection{The Definitions of Gap Parameters}\label{sec:rp-definition}

A planet-opened gap is a perturbation on top of an initial ``background'' disk with no planets. The definition and characterization of a gap are relative to this background. While in simulations a planetless background can be well defined, in real observations of actual systems it is nearly impossible to do so. In addition, a complete global disk profile is often not accessible in real observations with finite inner and outer working angles. In fact, based on scattered light alone it is usually difficult to verdict whether a ``gap-like'' structure is indeed a planet-opened gap or not. To mimic actual observations and to maximize the applicability of the correlations derived in this work, below we will assume gaps as local structures so that its parametrization only involves its immediate neighborhood, and we will characterize an observed gap in such a way that does not require prior knowledge of the underlying planetless background disk. We emphasize this is an necessary assumption in order to construct useful empirical relations to quickly and quantitatively convert observed gap properties to planet and disk properties. For individual actual systems, whether this assumption is good or not may be indicated by the reasonableness of the fitting. More accurate assessments of the planet and disk properties require detailed modeling of individual systems with specifically designed simulations to match the 1D (or even 2D) gap profiles, instead of using parameterized gap depth and width.

The definitions of gap depth and width in surface density vary somewhat in the literature. For the former, the common practice is to define it as the ratio between the minimum $\Sigma$ inside the gap (or averaged over a finite radial range) to a fiducial ``undepleted'' value ($\Sigma_0$). What counts as $\Sigma_0$ varies. For the gap width, the difficulty lies in defining the two gap edges, $r_{\rm out}$ and $r_{\rm in}$ (such that $\widthsigma=r_{\rm out}-r_{\rm in}$). The common practice is to define them as where $\Sigma$ reaches a certain threshold fraction of $\Sigma_0$ \citep[e.g.,][]{kanagawa16width}. This class of width definitions does not work for shallow gaps in which the depletion is less than the threshold fraction. With this fixed-depth-width definition, $\widthsigma$ depends on $\mplanet$ and $\alpha$ in addition to $h/r$ \citep[e.g.,][]{kanagawa16width, kanagawa16}.

In this paper, we adopt dynamical definitions of the gap depth and width in both surface density and images, as illustrated in Figure~\ref{fig:definition}. Our definitions work for of gaps with a wide range of depth and width. For $\Sigma$, we first find the location in the gap where $\Sigma$ reaches the minimum, $\rminsigma$. The initial value of $\Sigma$ at $\rminsigma$, $\Sigma_0(\rmin)$, is regarded as the fiducial undepleted value. Thus $\depthsigma$ is defined as 
\begin{equation}
\depthsigma\equiv\frac{\Sigma_0(\rminsigma)}{\Sigma(\rminsigma)}.
\label{eq:depthsigma}
\end{equation}
We then regard the inner and outer gap edges, $r_{\rm in,\Sigma}$ and $r_{\rm out,\Sigma}$, as where surface density reaches the geometric mean of the minimum and the undepleted value,
\begin{equation}
\Sigma_{\rm edge}=\Sigma(r_{\rm in,\Sigma})=\Sigma(r_{\rm out,\Sigma})= \sqrt{\Sigma_0(\rminsigma)\Sigma(\rminsigma)};
\label{eq:sigmaedge}
\end{equation}
therefore the gap width $\widthsigma$ is 
\begin{equation}
\widthsigma\equiv r_{\rm in,\Sigma}-r_{\rm out,\Sigma},
\label{eq:widthsigma}
\end{equation}
and the normalized gap width $\widthsigmanorm$ is defined as
\begin{equation}
\widthsigmanorm\equiv \frac{\widthsigma}{\rminsigma}.
\label{eq:widthsigmanorm}
\end{equation}
We note $\widthsigma$ has a physical unit (e.g., AU or arcsec), while $\widthsigmanorm$ is dimensionless.

The definitions of gap width and gaps in scattered light are similar. We first scale the image by $r^2$, then find the location in the gap where $I$ reaches the minimum, $\rminimage$ (a gap cannot be defined if the $r^2$-scaled intensity monotonically varies with radius). We find that in general $\rminimage\sim\rminsigma\sim\rp$, with small model-by-model differences. We then read the intensity at $r_1=2\rminimage/3$ and $r_2=3\rminimage/2$. These two locations are far away from the gap region so they are considered to be outside the influence of the gap. A fiducial ``background'' $I_0$ at $\rmin$ is subsequently defined as the geometric mean of the two points (note that this choice assumes the underlying ``planetless'' background profile follows a power law), $I_0(\rminimage)\equiv\sqrt{I(r_1)I(r_2)}$, and the gap depth is
\begin{equation}
\depthimage\equiv I_0(\rminimage)/I(\rminimage).
\label{eq:depthimage}
\end{equation}
The inner and outer edges of the gap, $r_{\rm in,I}$ and $r_{\rm in,I}$, are found at where $I$ reaches the geometric mean of $I_0$ and $I$ at $\rminimage$.
\begin{equation}
I_{\rm edge}=I(r_{\rm in,I})=I(r_{\rm out,I})= \sqrt{I_0(\rminimage)I(\rminimage)}.
\label{eq:imageedge}
\end{equation}
Correspondingly, the width of the gap $\widthimage$ is defined as
\begin{equation}
\widthimage\equiv r_{\rm in,I}-r_{\rm out,I},
\label{eq:imagewidth}
\end{equation}
and the normalized gap width $\widthimagenorm$ is
\begin{equation}
\widthimagenorm\equiv \frac{\widthimage}{\rminimage}.
\label{eq:widthimagenorm}
\end{equation}
The gap parameters in surface density and in face-on images for all models are listed in Table~\ref{tab:models}.

Finally, we note that all image-related quantities, such as $\depthimage$, $\widthimage$, and $\widthimagenorm$, are functions of both angular resolution $\eta$, and inclination $i$, i.e., $\depthimage(\eta,i)$, $\widthimage(\eta,i)$, and $\widthimagenorm(\eta,i)$. We omit the variables when the context is clear.

\subsection{The Correlations between $[\alpha, h/r, \mplanet]$ and $[\widthsigmanorm,\depthsigma]$}\label{sec:sigma}

Previous numerical studies have fit empirical relations to express $\widthsigma$ and $\depthsigma$ as functions of $[\alpha, h/r, \mplanet]$. Here we derive our version based on the new definitions of depth and width in Figure~\ref{fig:definition}. As discussed in Section~\ref{sec:intro}, analytical weakly-nonlinear theory and simple scaling relations based on torque balancing have motivated Equations~\ref{eq:gapdepthsigma} and \ref{eq:width-h} as possible functional forms for $\widthsigma$ and $\depthsigma$. We are able to achieve good fit for both correlations, as shown in Figure~\ref{fig:width_depth_sigma}, by varying only the constant proportionality factor. For gap width we have
\begin{equation}
\widthsigmanorm=5.8\frac{h}{r}.
\label{eq:width-sigma-h}
\end{equation}
The standard deviation $\sigma$ is $9\%$ for all models. If we restrict to only the models that have been run for longer than viscous timescale ($\alpha=10^{-3}$ and $\alpha=3\times10^{-3}$, the black points), we again recover Eqn~(\ref{eq:width-sigma-h}) while $\sigma$ shrinks to $6\%$. We note that if we adopt the \citet{kanagawa16width} definition of the gap width, we do recover their $\widthsigmanorm=0.41q^{1/2}(h/r)^{-3/4}\alpha^{-1/4}$ correlation (with a standard deviation of 17\%). However we prefer our dynamical definition of the gap width over defining the width as the distance between two points on the edges with a fixed depletion factor for a few reasons. First, the latter does not work for shallow gaps opened by relatively low mass planets, and in real observations such gaps are more common than deeper gaps opened by more massive planets, due to the higher occurrence rate of smaller planets than their more massive siblings \citep{cumming08}. Second, adopting the latter definition returns a less robust correlation between the gap width in images and the gap width in surface density (not shown). Third, while the \citet{kanagawa16width} correlation is a pure empirical correlation synthesized from simulations within a confined parameter space, our Eqn (\ref{eq:width-sigma-h}) has some theoretical footing from gap opening theory, thus more robust when applying to real systems that may lie outside the parameter space explored by our models.

For gap depth, we have
\begin{equation}
\depthsigma-1=0.043~q^2\left(\frac{h}{r}\right)^{-5}\alpha^{-1},
\label{eq:depth-sigma}
\end{equation}
with $\sigma=17\%$ for all but the three $\mplanet=\mj$ and $h/r=0.05$ models, which have the largest thermal masses in our models, $\mplanet=8M_{\rm thermal}$, and so their gap opening process is in the strongly nonlinear regime and the analytical torque balancing calculations break. We have experimented with freeing the power law indexes in the fit; overall no significant improvement is found.

We thus conclude that the gap width and depth definitions in Figure~\ref{fig:definition} successfully capture the physics of gap opening as argued by analytical theories. One thing to note is that $h/r$ can now be directly constrained based on measured gap properties. Assuming hydrostatic equilibrium in the vertical direction, scale height $h/r$ is related to disk temperature $T$ at each radius as
\begin{equation}
\frac{h}{r}=\frac{c_{\rm s}}{v_{\rm k}}=\frac{\sqrt{kT}/\mu}{v_{\rm k}},
\label{eq:h/r}
\end{equation}
where $v_{\rm k}$ is the Keplerian speed, and $\mu$ is the mean molecular weight. Thus, our constraints on $h/r$ based on measured gap properties can be used to infer the midplane temperature of the disk at the location of the gap.

\subsection{The Correlations between Gap Properties in Surface density and in Face-on Images}\label{sec:faceon}

At $i=0$, we measure 4 quantities for each model for a given $\eta$: $\widthsigmanorm$, $\depthsigma$, $\widthimagenorm(\eta)$, and $\depthimage(\eta)$. Figure~\ref{fig:width_depth} compares $\widthsigmanorm$ with $\widthimagenorm(\eta)$, and $\depthsigma$ with $\depthimage(\eta)$, for both $\eta=0$ and $\eta=0\arcsec.04$. 

In full resolution images, it is well approximated that $\widthimagenorm(0)=\widthsigmanorm$, with a standard deviation of 7\% for all gaps. For the depth, $\depthimage(0)$ is close to $\depthsigma$ for shallow gaps ($\depthsigma\lesssim5$), but deviates from $\depthsigma$ for deeper gaps. Once convolved, we expect the gap to become shallower and narrower, and we expect the effect of PSF smearing to depend on the ratio of 
\begin{equation}
\beta=\eta/\widthimage,
\label{eq:beta}
\end{equation}
since with higher angular resolution (smaller $\eta$) and/or wider gaps (bigger $\widthimage$), the intrinsic gap profile at $\eta=0$ is better preserved; and vice versa. This is confirmed in the figure --- for wider gaps $\widthimagenorm(\eta)$ well correlates with $\widthsigmanorm$ (as well as $\widthimagenorm(0)$), while narrower gaps becomes noticeably wider with finite $\eta$. For the depth, PSF smearing makes the gap shallower in all cases.

Quantitatively, we find $\widthimagenorm(\eta)$ is correlated to $\widthsigmanorm(\eta)$, $\beta$, and $\depthimage$ in the following way,
\begin{equation}
\widthimagenorm = \sqrt{ \widthsigmanorm^2 + 0.13\frac{\beta^2}{\depthimage}},
\label{eq:width-image-sigma}
\end{equation}
with $\sigma=9$\%. In the high angular resolution limit ($\beta\rightarrow0$), $\widthimagenorm\rightarrow\widthsigmanorm$. For the gap depth, we find
\begin{equation}
\depthimage = \left( \frac{\depthsigma}{1+0.0069\depthsigma} \right) ^{0.85-0.44\beta^2},
\label{eq:depth-image-sigma}
\end{equation}
with $\sigma=20$\%. In the high angular resolution ($\beta\rightarrow0$) and shallow gap limit ($\depthsigma\rightarrow1$), $\depthimage\rightarrow\depthsigma$. Both equations are invertible and it is straight forward to obtain $\depthsigma$ and $\widthsigma$, given $\depthimage$, $\widthimage$, and $\eta$. These correlations are synthesized for the parameter space of $0.2\lesssim\widthimagenorm\lesssim0.5$ and $1\lesssim\depthimage\lesssim50$; we caution their applications to the parameter space beyond.

\subsection{Effects of Inclinations}\label{sec:inclinations}

In this section, we compare the radial profile of face-on images to the radial profile along the major axis of disks at modest inclinations ($i\lesssim45^\circ$). In the latter, the radial profiles are averaged over a wedge of $\pm10^\circ$ from the major axis centered on the central star, as indicated in Figure~\ref{fig:image_rp_h5a1e3_45}(c). We note that in inclined disks, equal distance points from the center approximately fall on ellipses not centered on the star (shifted along the minor axis), because these ellipses are at non-zero height \citep[Fig. 1 in][]{stolker16}. Therefore, strictly speaking the major axis going through the star is not the major axis of the gap ellipse (the two are parallel to each other). Nevertheless, measuring the radial profile along the major axis centered on the star is common practice in the field \citep[e.g.,][]{ginski16, deboer16}; to facilitate the applications of our results, we do the same. The scattering angle along the major axis is close to $90^\circ$, the same as in face-on images. Therefore, the relative radial profile along the major axis is minimally affected by dust scattering properties.

Major axis radial profiles at modest inclinations closely follow their face-on counterparts. Figure~\ref{fig:image_rp_h5a1e3_45} compares the two for three representative models with $\alpha=10^{-3}$, $h/r=0.05$, and $\mplanet=2\mn$, $1\ms$, and $1\mj$. In all cases the gap region in the inclined disk is slightly shifted inward comparing with the face-on disk, due to the the inclination effect discussed above. The outer gap edge is shifted slightly more than the inner edge. Figure~\ref{fig:image_rp_1msh5a1e3_i} shows the gap profile at $i=0$, $15^\circ$, $30^\circ$, and $45^\circ$ for both the full resolution and convolved images of model 1$M_{\rm S}h$5$\alpha$1e-3. Intuitively, the differences between inclined and face-on images increase with $i$. When $i$ approaches 0, the differences diminish.

Quantitatively, the gap width in inclined disks agree with face-on disks very well, 
\begin{equation}
\widthimagenorm(45^\circ)=\widthimagenorm(0),
\label{eq:width-i}
\end{equation}
with a standard deviation of 5\% between the two. On the other hand, inclined images have systematically shallower gaps,
\begin{equation}
\depthimage(i)=\depthimage(0)(1-1.2\times10^{-6}i^3),
\label{eq:depth-i}
\end{equation}
with a standard deviation of 9\% ($i$ is in the unit of degree). The effect of inclination on $\depthimage(i)$ is weak --- with $i=45^\circ$, the gap depth is only reduced by 10\%. Figure~\ref{fig:depth_depth_45} compares $\widthimagenorm(45^\circ)$ {\it vs} $\widthimagenorm(0)$, and $\depthimage(45^\circ)$ {\it vs} $\depthimage(0)$ for 6 representative models.

\subsection{Effects of the Flareness of the Disk}\label{sec:flareness}

So far, $h/r$ is taken to be a constant in our models (i.e., no radial dependence). This choice results in a flat disk surface, which simplifies the problem in our qualitative analysis of the physical picture, and enables us to obtain a straightforward intuition to why scattered light features closely follow the structures in surface density (Section~\ref{sec:qualitative}). However, real disks passively heated by the central star may be flared, with $h/r\propto r^\gamma$ and $\gamma>0$ \citep[e.g.,][]{chiang97}. For example, by modeling the spectral energy distribution (SED) and resolved mm observations, \citet{andrews11} obtained $\gamma=0.19\pm0.1$ for a sample of 12 transitional disks. This surface flareness certainly affects the scattered light distribution in disks; for example, flared disks have surface brightness decreasing slower with radius than flat disks. This may affect the width and depth of gaps in images, and alter the surface density profiles slightly (through Eqns~\ref{eq:depth-sigma} and \ref{eq:width-sigma-h}).

However, the narrow gaps explored in this study are essentially local structures. In this case, the $h/r$ variation across the gap under a reasonable $\gamma$ (e.g., $\gamma\lesssim0.25$) may be too small to produce any significant effect. To test this hypothesis, we carry out experiments and compare three models with different $\gamma$'s, but otherwise the same parameters, to examine the dependence of gap width and depth on $\gamma$. The results are shown in Figure~\ref{fig:sigma_image_diffhrpower_rp}. With $h/r=0.05$ at $\rp=30$ AU, $\alpha=10^{-3}$, and $\mplanet=1\ms$, the three models with $\gamma=0$ (flat disk), 0.1, and 0.25 (flare disks) produce very similar gap properties in both the surface density and scattered light (the global scattered light profile in these models are of course different --- flared disks have a brighter outer disk --- as expected). Specifically, for $\widthsigmanorm$ and $\widthimagenorm$, all three models return essentially the same values; for $\depthsigma$, the two flared models have measurements higher than the flat model by $\sim5\%$; and for $\depthimage(\eta|0)$, the two flared models differ from the flat model by about 10\%. Therefore, we tentatively conclude that under reasonable flareness ($\gamma\lesssim0.25$), the flaring of the disk does not significantly affect the gap properties in both the density and image space for narrow gaps as in this example; subsequently, we expect the correlations derived in this paper to hold to a good approximation. A full parameter space exploration is needed to firmly establish this statement for more general cases.

\subsection{A Generic Roadmap to Link $[\widthimagenorm,\depthimage]$ with $[\mplanet,h/r,\alpha]$}\label{sec:roadmap}

Fitted correlations~\ref{eq:width-sigma-h}--\ref{eq:depth-i} compose a complete set of equations to link $[\widthimagenorm,\depthimage]$ with $[\mplanet,h/r,\alpha]$. Once a narrow gap ($\widthimagenorm(\eta,i)\lesssim0.5$) is identified in scattered light of a disk at a modest inclination ($i\lesssim45^\circ$) with both the inner and outer gap edges clearly revealed, one can follow the following steps to link observed gap properties to planet and disk parameters $[\mplanet,h/r,\alpha]$:
\begin{enumerate}[I]
\item Use Equations~\ref{eq:width-i}--\ref{eq:depth-i} to eliminate the effect of inclination: $\depthimage(\eta,i)\rightarrow\depthimage(\eta,0)$, $\widthimagenorm(\eta,i)\rightarrow\widthimagenorm(\eta,0)$;
\item Use Equations~\ref{eq:beta}--\ref{eq:depth-image-sigma} to eliminate the effect of finite angular resolution, and link images to surface density: $\depthimage(\eta,0)\rightarrow\depthsigma$, $\widthimagenorm(\eta,0)\rightarrow\widthsigmanorm$;
\item Finally, Equations~\ref{eq:width-sigma-h} and \ref{eq:depth-sigma} link the gap properties in surface density to $[\mplanet,h/r,\alpha]$. Note that normally $h/r$ can be constrained directly from the gap width, and it leaves the quantity $q^2/\alpha$ to be constrained from the gap depth.
\end{enumerate}

We note that these correlations are derived for polarized intensity at $H$-band, but they also apply to total intensity images, and/or images in other spectral bands within a factor of $\sim2$ difference in wavelength from $H$-band (as long as the signals are dust scattering dominated), because both depth and width are measured in a relative sense, and face-on radial profiles and major axis radial profiles in inclined images minimize the dependence on scattering angles. Finally, we emphasize that a key precondition when applying these correlations to actual systems is that the gap bottom needs to be detected (i.e., the gap bottom in images should be above the detection threshold). If not, the gap depth becomes illy defined.

\subsection{Cautions in Applying Our Results and Possible Future Improvements}\label{sec:caveat}

In simulating the gap images and deriving the correlations, we have made a number of assumptions in disk structures and modeling. Here we comment on the effects of these assumptions, and caution the readers about the caveats when applying our results to real disks.
\begin{enumerate}

\item \citet[][Figure 2]{jangcondell12} highlighted that the outer gap edge may receive extra stellar radiation due to the depletion of material inside the gap, leading to higher temperature thus higher $h/r$ (see also \citealt{jangcondell08, isella16}). However, whether this effect can increase the contrast of the gap at the outer edge need additional investigations. \citet[][Figure 6]{fung16} showed that gas inside the gap and around the gap edges circulate, therefore the heating at the outer edge is redistributed to a much larger region in the disk, weakening this thermal feedback. Future coupled hydro-radiation-transfer simulations are needed to quantify this effect.
\item \citet[see also \citealt{hernandez08}]{ribas14} showed that the fraction of stars with protoplanetary disks, indicated by their infrared excesses, drops on a time scale of about 3 Myr. The timescale that a planet takes to fully open a gap with a gap width of $6h$ (i.e., our gap width) is approximately the viscous timescale to cross $6h$, $\tau_\nu = 6 / ( \alpha (h/r) \Omega_{\rm K} )$. With $\alpha (h/r) / (\rp/ 30 \rm {AU})^{1.5} \lesssim 10^{-4}$, $\tau_\nu$ is longer than 3Myr. Therefore, in low viscosity disks, very young systems, and large separations from the center, gaps may not reach their final depth and width, which invalids the $[\mplanet,h/r,\alpha]-[\widthsigmanorm,\depthsigma]$ correlations synthesized in this work and in others. This gap opening timescale issue is however irrelevant to the $[\depthimage,\widthimagenorm]-[\depthsigma,\widthsigma]$ conversions, and we expect them to hold even when the gap is only partially opened.
\item In our MCRT simulations, no noise is added into the images (apart from the intrinsic Poisson noise due to finite number of photons), thus the gap bottom reach their theoretically minimum. If in real observations the gap bottom reaches the noise level, the measured gap depth, and the derived planet mass, will be their lower limits, while the measured gap width, and the derived disk scale height, will be their upper limits.
\end{enumerate}


\section{Applications to the Gaps in HD 169142, TW Hya, HD 97048, LkCa~15, and RX J1615.3-3255}\label{sec:applications}

As examples, in this section we apply our results to a few gaps imaged in actual systems (Figure~\ref{fig:real_disks}; ``gaps'' in these systems refer to the regions around local minimums on the $r^2$-scaled scattered light profile). The gap bottom in these disks are robustly detected. Assuming each gap is opened by a single planet, we derive $h/r$ and $q^2/\alpha$ at the gap location, as well as the planet mass for several assumed $\alpha$. The results are summarized in Table~\ref{tab:applications}. We stress that the derived disk and planet properties in the table should be taken as suggestive values only, as real disks may not obey the assumptions in our models outlined in Section~\ref{sec:caveat}, and some of the gaps observed so far may or may not be opened by (a single) planet. Actual hydro+MCRT modeling of individual disks with planet-disk interaction models quantitatively fitting the observations is encouraged to more accurately recover the disk and planet properties in specific systems. 


\subsection{HD 97048}\label{sec:hd97048}

HD 97048 is a $\sim2.5\msun$ Herbig Ae/Be star located at $\sim$158 pc \citep{vandenancker98, vanleeuwen07} surrounded by a protoplanetary disk several hundreds of AU in size. The disk is inclined at $\sim40^\circ$ based on mm observations \citep{walsh16, vanderplas16}. \citet{maaskant13} first suggested that this disk may be gap/cavity harboring based on SED modeling. Very recently, VLT/SPHERE $J$-band imaging in both polarized and total intensity found multiple rings and gaps in this system \citep{ginski16}. In addition, the mm dust continuum counterparts of some of these structures may have been found by \citet{vanderplas16}, suggesting the scattered light gaps and rings are likely to be physical density structures instead of shadow effects. 

Here we focus on the ``gap 2'' in the $J$-band VLT/SPHERE dataset \citep{ginski16}. Comparing with the other gaps, gap 2 is well detected in both polarized and total intensity, with inner and outer edges clearly resolved. We obtain $h/r=0.06$ at the location of the gap ($0\arcsec.67$; 106 AU), and constrain the planet mass to be between 0.4--4 $\mj$ for $10^{-4}\lesssim\alpha\lesssim10^{-2}$. By analyzing the offset of the center of the gap ellipse from the star, \citet{ginski16} determined that at the gap location the disk's surface is at $\theta\sim0.2$ rad. Therefore, the disk surface is at about 3--4 scale heights away from the disk midplane, which is reasonable \citep[e.g.,][]{chiang97} (we note that in our MCRT simulations the disk surface is also generally about 4 scale heights away from the midplane). \citet{ginski16} ruled out the presence of planets more massive than $2\mj$ at the gap location (assuming the BT-SETTL isochrones; \citealt{allard11}). Comparing with our constraints on the planet mass, this suggests the viscosity in the gap may be low, $\alpha\lesssim10^{-3}$. 

\subsection{TW Hya}\label{sec:twhya}

TW Hya is a $\sim0.8\msun$ K6 star located at 54 pc \citep{torres06, torres08}. The nearly face-on disk ($i\sim7^\circ$; \citealt{qi04}) has been imaged by HST \citep{debes13, debes16}, Gemini/GPI \citep{rapson15twhya}, Subaru/HiCIAO \citep{akiyama15}, and VLT/SPHERE \citep{vanboekel16}, and multiple gaps have been identified in scattered light: one at $\sim$80 AU, one at $\sim20$ AU, and one at $\lesssim6$ AU. The mm continuum counterpart of the 20 AU gap may have been identified by \citet{andrews16} in ALMA observations at a slightly larger radius (possibly caused by the dust filtration effect).

Here we examine two gaps in the system: the one at $\sim20$, and the one at $\sim80$ AU, in the $H$-band VLT/SPHERE dataset \citet[Fig. 3]{vanboekel16}. Due to the low inclination, the azimuthally averaged radial profile is used. For the inner gap, $h/r$ at the gap location ($0\arcsec.37$; 20 AU) is 0.068, and the planet mass is between 0.05--0.5 $\mj$ for $10^{-4}\lesssim\alpha\lesssim10^{-2}$; for the outer gap, $h/r$ at the gap location ($1\arcsec.5$; 81 AU) is 0.055, and the planet mass is between 0.03--0.3 $\mj$ for $10^{-4}\lesssim\alpha\lesssim10^{-2}$. 

Using hydro+MCRT simulations, \citet{rapson15twhya} tentatively fit the observed gap profile at $\sim20$ AU in the Gemini/GPI dataset using a $0.16\mj$ planet at 21 AU assuming $\alpha=10^{-3}$ and $h/r=0.068$. Their results agree well with our findings. \citet{vanboekel16} used radiative transfer simulations to transform the scattered light map into surface density and scale height profiles. At 20 AU they obtained $h/r=0.05$, while at $80$ AU they obtained $h/r=0.08$. Subsequently, assuming $\alpha=2\times10^{-4}$, they derived the masses of the planets at 20 and 80 AU to be 0.05$\mj$ and 0.11$\mj$, based on the a similar version of Eqn~\ref{eq:depth-sigma} in \citet{duffell15gap}. While their $h/r$ and $\mplanet$ at the inner gap is broadly consistent with our results, our estimated $h/r$ for the outer gap is significantly lower (and also lower than our estimated $h/r$ at 20 AU), resulting in a lower estimate for $\mplanet$ as well. This may indicate that the 80 AU gap is not opened by a planet; alternatively, this may be because our assumption of the gap structure being local is no longer valid for the 80 AU gap, which may lie in the shadow created by the inner disk thus the underlying ``background'' is no longer smooth.

\subsection{HD 169142}\label{sec:hd169142}

HD 169142 is a $\sim2\msun$ Herbig Ae star located at 145 pc \citep{raman06, sylvester96}. It has a prominent protoplanetary disk at an inclination of $13^\circ$, determined from gas kinematics \citep{raman06, panic08}. A narrow gap around 50 AU was first discovered in $H$-band VLT/NACO polarized light imaging \citep{quanz13gap}, and subsequently found by Subaru/HiCIAO \citep{momose15} and in dust thermal emission at 7mm by VLA \citep{osorio14}. The radius-varying disk profile was interpreted as two power laws in the inner and outer disks joined by a transition region in between by \citet{momose15}. \citet{osorio14} also discovered a blob in the 7mm VLA dataset residing right inside the gap, with an estimated total mass of $6\times10^{-6}\msun$ in dust. Through radiative transfer modeling, \citet{wagner15169142} determined that this gap cannot be explained by shadow effects caused by the inner disk, leaving the planet-sculpting scenario as a favorite hypothesis. By comparing the apparent gap depth in 7mm observations with their $\mplanet-\depthsigma$ relation, \citet{kanagawa15} estimated the mass of the putative planet to be $\mplanet\gtrsim0.4\mj$. As \citet{rosotti16} pointed out, such estimate is risky as mm observations trace mm-sized grains, which can have substantially different spatial distribution from the gas due to dust/gas coupling effects. The disk has another inner gap at $\sim25$ AU \citep{honda12, quanz13gap}, and a companion candidate at the edge of the inner gap \citep[][see also \citealt{biller14}]{reggiani14}.

Here we examine the gap at 40--70 AU in the Subaru/HiCIAO $H$-band dataset \citep{momose15}. Due to the low inclination, we adopt azimuthally averaged radial profile after deprojecting the disk \citet[Fig. 2; assuming $i=13^\circ$ and position angle = 5$^\circ$]{momose15}. We derive $h/r$ at the gap location ($0\arcsec.35$; 51 AU) to be 0.08, and the planet mass is between 0.2--2.1 $\mj$ for $10^{-4}\lesssim\alpha\lesssim10^{-2}$.

\subsection{LkCa~15}\label{sec:LkCa15}

LkCa~15 is a $\sim1\msun$ K3 star located at 140 pc \citep{pietu07, simon00}. Using Spitzer IRS spectrum modeling, \citet{espaillat10} identified the system as a transitional disk with an outer disk, an inner disk, and a gap in between. SMA mm observations resolved the gap and determined its outer radius to be 50 AU in dust continuum emission \citep{andrews11}. The gap has since been resolved in scattered light by Subaru/HiCIAO \citep{thalmann10}, Gemini/NICI \citep{thalmann14}, and VLT/SPHERE \citep{thalmann15, thalmann16} in both total and polarized intensity. The latest VLT/SPHERE observations by \citet{thalmann16} at multiple optical to NIR bands clearly showed that the system has an substantial inner disk, and the gap is narrow enough ($\widthimagenorm<0.5$) to warrant the hypothesis that it may be opened by a single planet. The inclination of the disk is $\sim50^\circ$ based on mm observations \citep{andrews11, pietu07, vandermarel15}.

Here we examine the gap profile along the major axis of the disk in the VLT/SPHERE $H$-band polarized intensity dataset presented in \citet{thalmann16}. We derive $h/r$ at the gap location ($0\arcsec.26$; 36 AU) to be 0.07, and the planet mass is between 0.15--1.5 $\mj$ for $10^{-4}\lesssim\alpha\lesssim10^{-2}$.

LkCa has several detected planet candidates. Using non-redundant aperture masking interferometry on Keck, \citet{kraus12} identified a point source at a deprojected distance of $\sim20$ AU from the star. Recently, \citet{sallum15} reported the detections of two additional point sources in the system with LBT/LBTI and Magellan/MagAO. The three planet candidates in \citet{sallum15} are located between 15--19 AU. We note that these planet candidates are unlikely to be the one responsible for maintaining the gap edge in scattered light at $\sim50$ AU (i.e., the one whose properties we are inferring here), as they are too far away from the gap edge. The radial profile of the gap (Figure~\ref{fig:real_disks}d) is not inconsistent with the gap being opened by just one planet. Additional planets at $r\lesssim20$~AU with sufficiently low mass will not significantly affect the gap opened by the outer planet.

\subsection{RX J1615.3-3255}\label{sec:j1615}

RX J1615.3-3255 (hereafter J1615) is a $1.1\msun$ K5 star at 185 pc \citep{wichmann97, wahhaj10, andrews11}. The system has a gap with an outer radius of $\sim20-30$ AU, revealed in mm dust continuum emission first by SMA \citep{andrews11} and subsequently by ALMA \citep{vandermarel15}. The disk has a modest inclination, $i\sim45^\circ$. Recently, \citet{deboer16} resolved this system in scattered light at multiple optical-to-NIR wavelengths using VLT/SPHERE, and identified multiple gaps and rings in the system.

Here we examine the the major axis profile of the gap at $\sim90$ AU in the $J$-band VLT/SPHERE dataset presented by \citet[marked ``G'' in Fig. 1]{deboer16}. We derive $h/r$ at the gap location ($0\arcsec.52$; 96 AU) to be 0.06, and the planet mass is between 0.07--0.7 $\mj$ for $10^{-4}\lesssim\alpha\lesssim10^{-2}$.

By analyzing the alternating bright/dark pattern on the rings, \citet{deboer16} found tentative evidence to suggest that the rings in J1615 might be caused by shadows (i.e., variation in scale height instead of surface density). At the moment the evidence is inclusive. 


\section{Summary}\label{sec:summary}

Combing 2D and 3D hydro simulations with 3D radiative transfer simulations, we examine the morphology of planet-opened gaps in near-infrared scattered light images. Quantitatively, we obtain correlations between gap depth and width in inclined disks with finite angular resolution, to the intrinsic gap depth and width in face-on images with infinite resolution. The latter is subsequently quantitatively linked to the gap depth and width in disk surface density assuming parametrized $h/r$ profile across the gap region, which can be used to constraints the mass of the gap-opening planet mass $\mplanet$, the disk scale height at the location of the gap $h/r$, and disk viscosity $\alpha$. The main take aways are:
\begin{enumerate}
\item 2D hydro simulations, puffed up using an assumed midplane temperature profile, produce the same gap profile in scattered light images as 3D simulations.
\item With our definition illustrated in Figure~\ref{fig:definition}, the aspect ratio $h/r$ in the gap region can be directly backed out from the gap width. This can be used to constrain the midplane temperature in disks.
\item Equations~\ref{eq:width-sigma-h}--\ref{eq:depth-i} compose a complete set of correlations to link observed $[\widthimagenorm,\depthimage]$ to $[\mplanet,h/r,\alpha]$ for narrow gaps ($\widthimagenorm\lesssim0.5$) in disks with modest inclinations ($i\lesssim45^\circ$) and flareness ($h/r\propto r^{\lesssim0.25}$). Once such a gap is identified in scattered light with both the inner and outer gap edges clearly revealed, one can follow the steps outlined in Section~\ref{sec:roadmap} to constrain fundamental planet and disk parameters $\mplanet$, $h/r$, and $\alpha$.
\item We apply our results to the gaps imaged in scattered light in HD 97048, TW Hya, HD 169142, LkCa~15, and RX J1615.3-3255, to derive $h/r$ and $\mplanet^2/\alpha$ at the locations of their gaps. The results are listed in Table~\ref{tab:applications} (see also Figure~\ref{fig:real_disks}). Assuming $\alpha=10^{-3}$, the masses of all gap-opening planets are roughly between 0.1--1 $\mj$.
\end{enumerate}


\section*{Acknowledgments}

We thank Eugene Chiang, Roman Rafikov, Sascha Quanz, Avenhaus Henning, and Tomas Stolker for motivating this work, Munetake Momose for sharing with us the SEEDS images of HD 169142, Sascha Quanz for sharing with us the VLT/NACO image of HD 169142, Joel Kastner and Valerie Rapson for sharing with us the Gemini/GPI image of TW Hya, Christian Ginski and Tomas Stolker for sharing with us the VLT/SPHERE data of HD 97048, Christian Thalmann for sharing with us the VLT/SPHERE LkCa 15 data on HD 97048, Jos de Boer for sharing with us the VLT/SPHERE data on RX J1615.3-3255, and Andrew Youdin for insightful discussions. We also thank the referee, Takayuki Muto, for constructive suggestions that largely improved the quality of the paper. This project is supported by NASA through Hubble Fellowship grant HST-HF-51320.01-A (R. D.) awarded by the Space Telescope Science Institute, which is operated by the Association of Universities for Research in Astronomy, Inc., for NASA, under contract NAS 5-26555. J.F. gratefully acknowledges support from the Natural Sciences and Engineering Research Council of Canada, the Center for Integrative Planetary Science at the University of California, Berkeley, and the Sagan Fellowship Program funded by NASA under contract with the Jet Propulsion Laboratory (JPL) and executed by the NASA Exoplanet Science Institution.



\clearpage

\begin{sidewaystable}
\scriptsize
\centering
\begin{tabular}{c|ccc|ccc|ccc|ccc}
\hline
Model                 & $q$                & $h/r$ & $\alpha$         & $\depthsigma$ & $\widthsigma$ & $\widthsigmanorm$ & $\depthimage(\eta|0)$ & $\widthimage(\eta|0)$ & $\widthimagenorm(\eta|0)$ & $\depthimage(\eta|0\arcsec.04)$ & $\widthimage(\eta|0\arcsec.04)$ & $\widthimagenorm(\eta|0\arcsec.04)$ \\
                      &                    &       &                  &               & AU            &                   &               & arcsec        &                   &                            & arcsec                     &                                \\ \hline
1$\mj h$5$\alpha$3e-4 & $10^{-3}$          & 0.05  & $3\times10^{-4}$ & 803.5         & 11.5          & 0.36              & 49.0          & 0.081         & 0.41              & 31.5                       & 0.072                      & 0.34                           \\
1$\mj h$5$\alpha$1e-3 & $10^{-3}$          & 0.05  & $10^{-3}$        & 761.4         & 9.7           & 0.32              & 59.5          & 0.068         & 0.33              & 26.2                       & 0.064                      & 0.30                           \\
1$\mj h$5$\alpha$3e-3 & $10^{-3}$          & 0.05  & $3\times10^{-3}$ & 74.9          & 9.8           & 0.32              & 24.2          & 0.066         & 0.31              & 13.9                       & 0.063                      & 0.30                           \\
1$\mj h$7$\alpha$3e-4 & $10^{-3}$          & 0.07  & $3\times10^{-4}$ & 107.1         & 11.8          & 0.38              & 23.7          & 0.077         & 0.35              & 17.6                       & 0.071                      & 0.33                           \\
1$\mj h$7$\alpha$1e-3 & $10^{-3}$          & 0.07  & $10^{-3}$        & 25.7          & 11.8          & 0.39              & 12.1          & 0.074         & 0.35              & 8.6                        & 0.071                      & 0.34                           \\
1$\mj h$7$\alpha$3e-3 & $10^{-3}$          & 0.07  & $3\times10^{-3}$ & 7.6           & 12.0          & 0.39              & 4.6           & 0.076         & 0.36              & 4.0                        & 0.075                      & 0.36                           \\
1$\ms h$4$\alpha$3e-4 & $2.9\times10^{-4}$ & 0.04  & $3\times10^{-4}$ & 78.0          & 8.0           & 0.25              & 30.6          & 0.056         & 0.28              & 12.9                       & 0.056                      & 0.27                           \\
1$\ms h$4$\alpha$1e-3 & $2.9\times10^{-4}$ & 0.04  & $10^{-3}$        & 42.2          & 7.4           & 0.24              & 18.6          & 0.049         & 0.23              & 8.2                        & 0.056                      & 0.26                           \\
1$\ms h$4$\alpha$3e-3 & $2.9\times10^{-4}$ & 0.04  & $3\times10^{-3}$ & 9.9           & 7.7           & 0.25              & 6.4           & 0.052         & 0.25              & 4.2                        & 0.060                      & 0.28                           \\
1$\ms h$5$\alpha$3e-4 & $2.9\times10^{-4}$ & 0.05  & $3\times10^{-4}$ & 30.9          & 8.4           & 0.28              & 17.7          & 0.056         & 0.28              & 8.6                        & 0.057                      & 0.27                           \\
1$\ms h$5$\alpha$1e-3 & $2.9\times10^{-4}$ & 0.05  & $10^{-3}$        & 11.2          & 8.5           & 0.28              & 6.8           & 0.057         & 0.27              & 4.7                        & 0.060                      & 0.29                           \\
1$\ms h$5$\alpha$3e-3 & $2.9\times10^{-4}$ & 0.05  & $3\times10^{-3}$ & 4.1           & 8.8           & 0.29              & 3.4           & 0.057         & 0.27              & 2.6                        & 0.063                      & 0.29                           \\
1$\ms h$7$\alpha$3e-4 & $2.9\times10^{-4}$ & 0.07  & $3\times10^{-4}$ & 6.3           & 11.1          & 0.36              & 4.3           & 0.073         & 0.34              & 3.7                        & 0.073                      & 0.35                           \\
1$\ms h$7$\alpha$1e-3 & $2.9\times10^{-4}$ & 0.07  & $10^{-3}$        & 3.1           & 11.6          & 0.38              & 2.3           & 0.071         & 0.34              & 2.2                        & 0.075                      & 0.36                           \\
1$\ms h$7$\alpha$3e-3 & $2.9\times10^{-4}$ & 0.07  & $3\times10^{-3}$ & 1.7           & 11.8          & 0.37              & 1.6           & 0.071         & 0.34              & 1.4                        & 0.081                      & 0.37                           \\
2$\mn h$4$\alpha$3e-4 & $10^{-4}$          & 0.04  & $3\times10^{-4}$ & 10.6          & 6.3           & 0.21              & 7.8           & 0.045         & 0.22              & 4.1                        & 0.052                      & 0.25                           \\
2$\mn h$4$\alpha$1e-3 & $10^{-4}$          & 0.04  & $10^{-3}$        & 4.7           & 6.6           & 0.22              & 4.1           & 0.045         & 0.21              & 2.7                        & 0.057                      & 0.27                           \\
2$\mn h$4$\alpha$3e-3 & $10^{-4}$          & 0.04  & $3\times10^{-3}$ & 2.4           & 6.9           & 0.22              & 2.2           & 0.045         & 0.21              & 1.8                        & 0.061                      & 0.28                           \\
2$\mn h$5$\alpha$3e-4 & $10^{-4}$          & 0.05  & $3\times10^{-4}$ & 4.5           & 7.7           & 0.26              & 4.0           & 0.055         & 0.26              & 3.0                        & 0.059                      & 0.28                           \\
2$\mn h$5$\alpha$1e-3 & $10^{-4}$          & 0.05  & $10^{-3}$        & 2.5           & 8.2           & 0.27              & 2.1           & 0.058         & 0.28              & 1.9                        & 0.066                      & 0.31                           \\
2$\mn h$5$\alpha$3e-3 & $10^{-4}$          & 0.05  & $3\times10^{-3}$ & 1.6           & 8.8           & 0.28              & 1.5           & 0.053         & 0.25              & N/A\tablenotemark{(a)}                        & N/A                        & N/A                            \\
2$\mn h$7$\alpha$3e-4 & $10^{-4}$          & 0.07  & $3\times10^{-4}$ & 1.6           & 13.3          & 0.43              & 1.5           & 0.085         & 0.41              & 1.4                        & 0.086                      & 0.39                           \\
1$\mn h$4$\alpha$3e-4 & $5.1\times10^{-5}$ & 0.04  & $3\times10^{-4}$ & 3.5           & 6.1           & 0.20              & 3.2           & 0.044         & 0.20              & 2.4                        & 0.053                      & 0.25                           \\
1$\mn h$4$\alpha$1e-3 & $5.1\times10^{-5}$ & 0.04  & $10^{-3}$        & 2.1           & 6.5           & 0.21              & 1.9           & 0.046         & 0.21              & 1.6                        & 0.061                      & 0.28                           \\
1$\mn h$4$\alpha$3e-3 & $5.1\times10^{-5}$ & 0.04  & $3\times10^{-3}$ & 1.4           & 7.0           & 0.22              & 1.3           & 0.051         & 0.23              & N/A                        & N/A                        & N/A                            \\ \hline
\end{tabular}
\caption{Model parameters and gap properties in surface density and in face-on images. See the definition of parameters in Section~\ref{sec:rp-definition}. $^{(a)}$ Gap undefined by our definitions (Figure~\ref{fig:definition}).}
\label{tab:models}
\end{sidewaystable}

\begin{sidewaystable}
\scriptsize
\centering
\begin{tabular}{c|c|c|c|c|c|c}
\hline
                        & HD 97048\tablenotemark{(a)}             & TW Hya\tablenotemark{(b)}                & TW Hya\tablenotemark{(c)}                & HD 169142\tablenotemark{(d)}            & LkCa15        & RX J1615\tablenotemark{(e)}           \\
                                  & \citet{ginski16} & \citet{vanboekel16} & \citet{vanboekel16} & \citet{momose15} & \citet{thalmann16} & \citet{deboer16} \\ \hline
Band                              & $J$                  & $H$                   & $H$                   & $H$                  & $J$                  & $J$                \\
Radial Profile\tablenotemark{(f)}                               & Major Axis               & Azim-Averaged                 & Azim-Averaged        & Azim-Averaged         & Major Axis            & Major Axis        \\
$d$                               & 158 pc               & 54 pc                 & 54 pc                 & 145 pc               & 140 pc               & 185 pc             \\
$i$                               & $40^\circ$           & $7^\circ$             & $7^\circ$             & $13^\circ$           & $50^\circ$           & $45^\circ$         \\
$M_\star$ ($M_\odot$)             & 2.5                  & 0.8                   & 0.8                   & 2                    & 1                    & 1.1                \\
$\rminimage$             & 0$\arcsec$.67 (106 AU)                 & 0$\arcsec$.37 (20 AU)               & 1$\arcsec$.5 (81 AU)               & 0$\arcsec$.37 (54 AU)                 & 0$\arcsec$.26 (36 AU)              & 0$\arcsec$.52 (97 AU)           \\
$\widthimage$           & 0$\arcsec$.22                 & 0$\arcsec$.15            & 0$\arcsec$.48            & 0$\arcsec$.17                 & 0$\arcsec$.10           & 0$\arcsec$.17         \\
$\widthimagenorm$                 & 0.33                 & 0.41                  & 0.32                  & 0.47                 & 0.41          & 0.34        \\
$\depthimage$                     & 10.4                 & 1.7             & 1.7             & 2.1                  & 3.6             & 2.5          \\
$\eta$\tablenotemark{(g)}                  & 0$\arcsec$.04                 & 0$\arcsec$.05                  & 0$\arcsec$.05                  & 0$\arcsec$.06                 & 0$\arcsec$.05                & 0$\arcsec$.04              \\
$\beta$                           & 0.18                 & 0.33                  & 0.10                  & 0.35                 & 0.48                 & 0.23               \\ \hline
$\widthimagenorm(i|0)$            & 0.33                 & 0.41                  & 0.32                  & 0.47                 & 0.41                 & 0.34               \\
$\depthimage(i|0)$                & 11.3                 & 1.7                   & 1.7                   & 2.1                  & 4.2                  & 2.8                \\
$\widthsigmanorm$                 & 0.33                 & 0.40                  & 0.32                  & 0.46                 & 0.40                 & 0.34               \\
$\depthsigma$                     & 21                 & 2.0                   & 1.9                   & 2.5                  & 7.2                  & 3.5                \\ \hline
$h/r$\tablenotemark{(h)}                             & 0.056                & 0.068                 & 0.055                 & 0.079                & 0.069                & 0.058              \\
$q^2/\alpha$                      & 2.6$\times10^{-4}$             & 3.6$\times10^{-5}$              & 1.1$\times10^{-5}$              & 1.1$\times10^{-4}$             & 2.2$\times10^{-4}$             & 3.8$\times10^{-5}$           \\
$\mplanet/\mj$ ($\alpha=10^{-2}$)\tablenotemark{(i)} & 4.0                 & 0.48                  & 0.26                  & 2.1                 & 1.5                 & 0.68               \\
$\mplanet/\mj$ ($\alpha=10^{-3}$) & 1.3                 & 0.15                  & 0.08                  & 0.67                 & 0.47                 & 0.22               \\
$\mplanet/\mj$ ($\alpha=10^{-4}$) & 0.40                 & 0.05                  & 0.03                  & 0.21                 & 0.15                 & 0.07               \\ \hline
\end{tabular}
\caption{Applying our results to six gaps detected in scattered light. The radial profiles of the gaps are shown in Figure~\ref{fig:real_disks}. $^{(a)}$ The ``gap 2'' in \citet[Fig. 2]{ginski16}. $^{(b)}$ The $\sim21$ gap in \citet{vanboekel16}. $^{(c)}$ The $\sim85$ gap in \citet{vanboekel16}. $^{(d)}$ The 40--70 AU gap in \citet{momose15}. $^{(e)}$ The gap structure ``G'' in \citet[Fig. 1]{deboer16}. \tablenotemark{(f)} The source of the radial profile. For HD 97048 we use the major axis radial profile on the north side, as on the south side the polarized intensity inside the gap may be noise dominated \citep[Fig. 7]{ginski16}. For TW Hya, as the disk is close to face-on, we machine-read the azimuthally averaged $H$-band gap profile in \citet[Fig. 3]{vanboekel16}. For HD 169142, as the disk is close to face-on, we use azimuthally averaged radial profile after deprojecting the disk as in \citet[Fig. 2; assuming $i=13^\circ$ and position angle = 5$^\circ$]{momose15}. For LkCa~15 and J1615 we use the major axis radial profile averaged over the two sides in \citet[Fig. 4a; Deep]{thalmann16} and \citet[Fig. 7]{deboer16}. $^{(g)}$ Angular resolution of the observation as the FWHM of the PSF in the actual data. $^{(h)}$ Derived $h/r$ at the location of the gap. $^{(i)}$ Planet mass assuming the disk viscosity in the parenthesis. See Section~\ref{sec:applications} for details.}
\label{tab:applications}
\end{sidewaystable}

\begin{figure}
\begin{center}
\includegraphics[trim=0 0 0 0, clip,width=0.6\textwidth,angle=0]{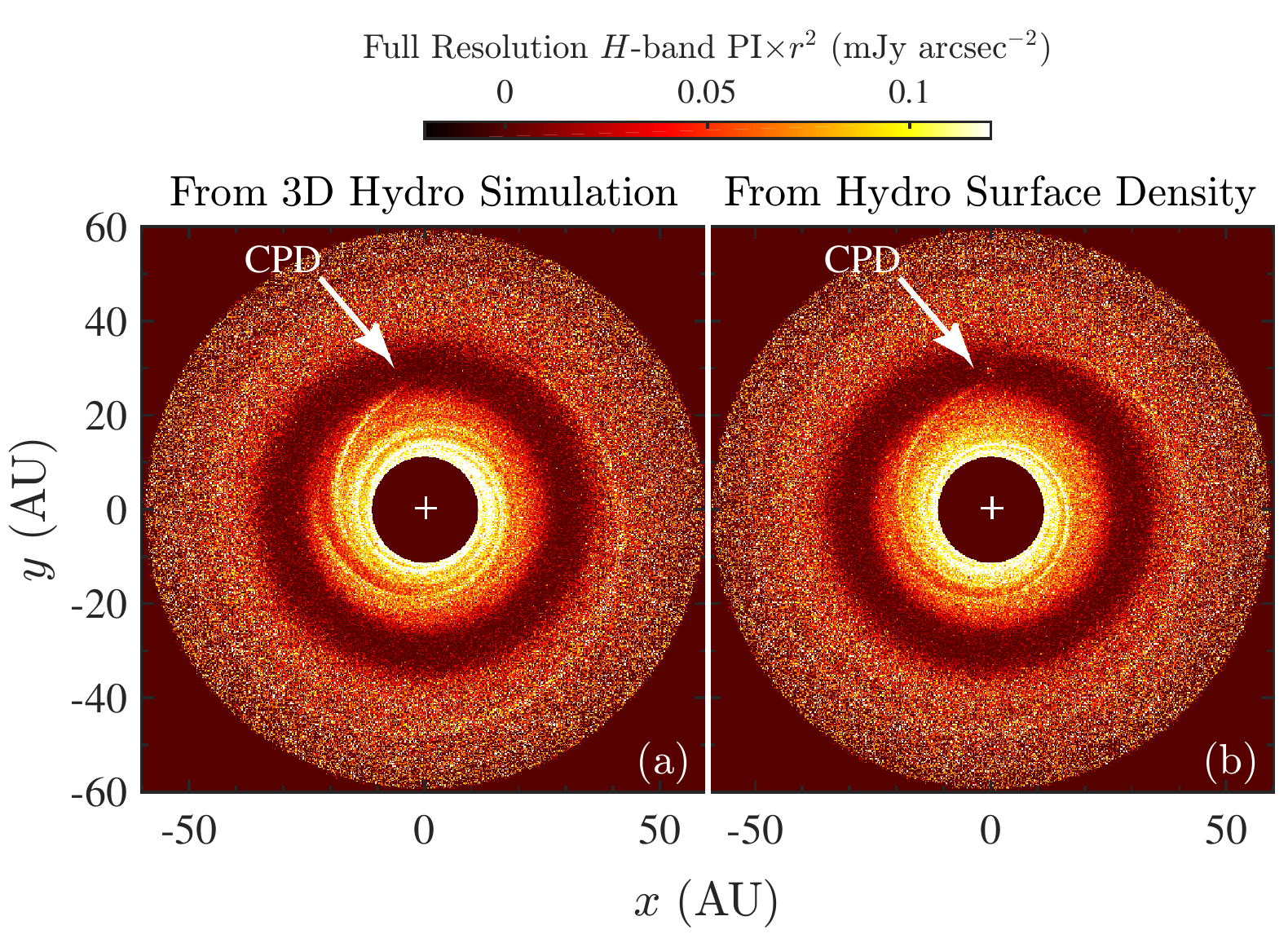}
\includegraphics[trim=0 0 0 0, clip,width=0.35\textwidth,angle=0]{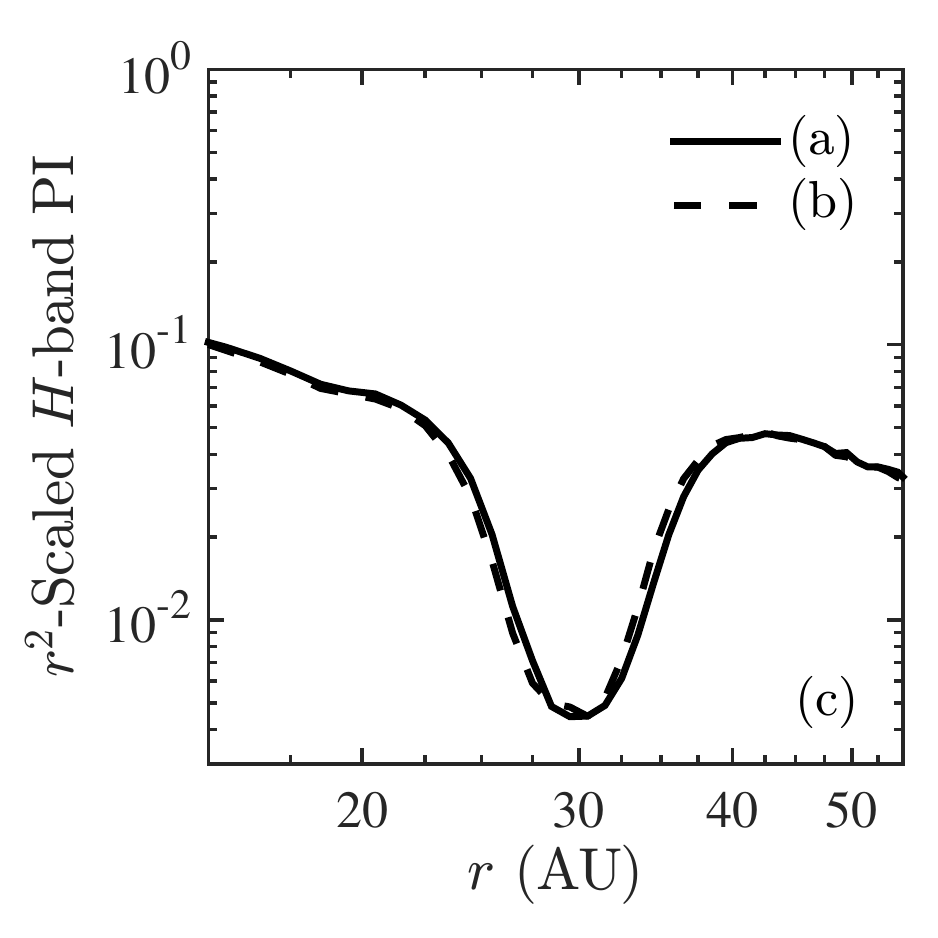}
\end{center}
\figcaption{{\bf Left:} Face-on images of a disk with $h/r=0.05$, $\alpha=10^{-3}$, and a $0.5\mj$ planet at 30 AU. Image (a) is from a 3D hydro simulation, while image (b) is produced by puffing up the surface density of the model by the same $h/r$ profile as in the 3D hydro calculation. Note that in both panels the circumplanetary region (CPD) is not excised (indicated by the arrow). {\bf Panel (c):} radial profile of the two images. The gap in the two images are essentially identical. See Section~\ref{sec:2d3d} for details.
\label{fig:2d3d}}
\end{figure}

\begin{figure}
\begin{center}
\includegraphics[trim=0 0 0 0, clip,width=\textwidth,angle=0]{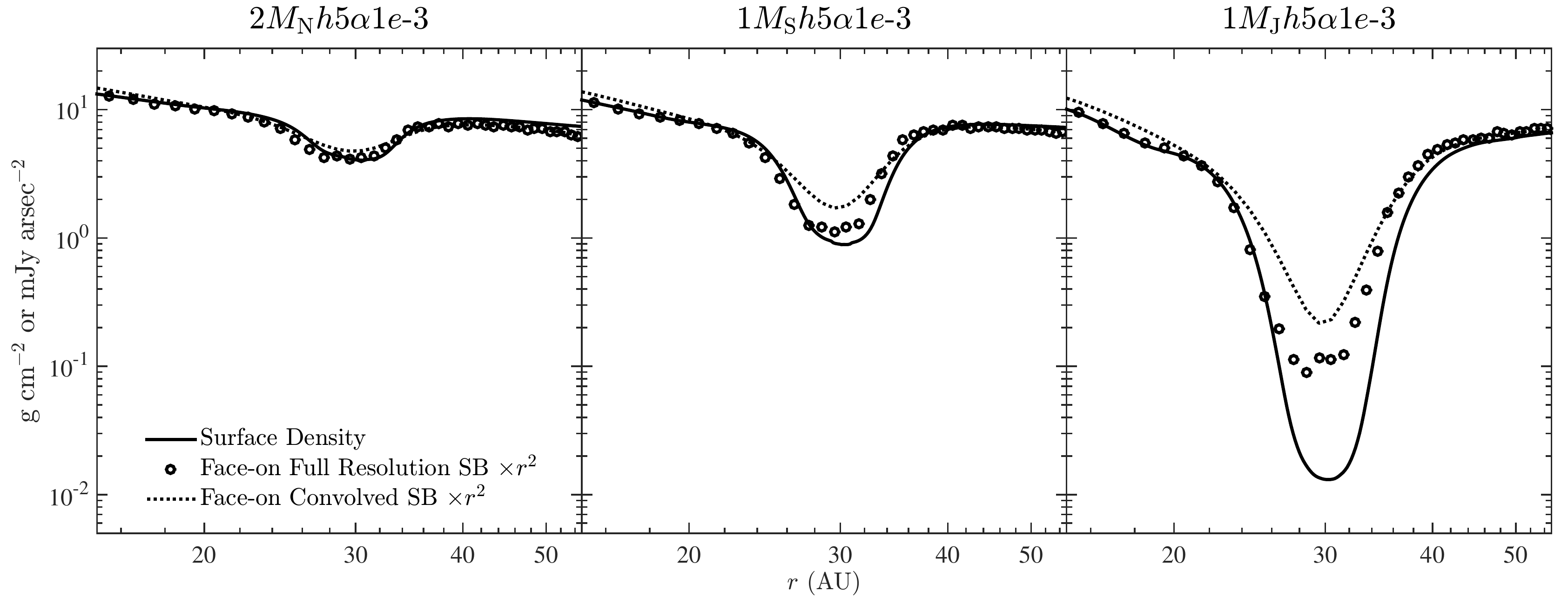}
\end{center}
\figcaption{Radial profiles of the surface density (solid) and $H$-band polarized intensity surface brightness (circles: full resolution; dotted line: convolved; both scaled by $r^2$) of three models with $h/r=0.05$, $\alpha=10^{-3}$, and $\mplanet=2\mn$, $1\ms$, and $1\mj$, {\bf from left to right}. Both the full resolution and convolved surface brightness curves have been scaled by the same constant factor so that they roughly meet the surface density curves at $r=15$ AU.
\label{fig:sigma_image_rp}}
\end{figure}

\begin{figure}
\begin{center}
\includegraphics[trim=0 0 0 0, clip,width=0.49\textwidth,angle=0]{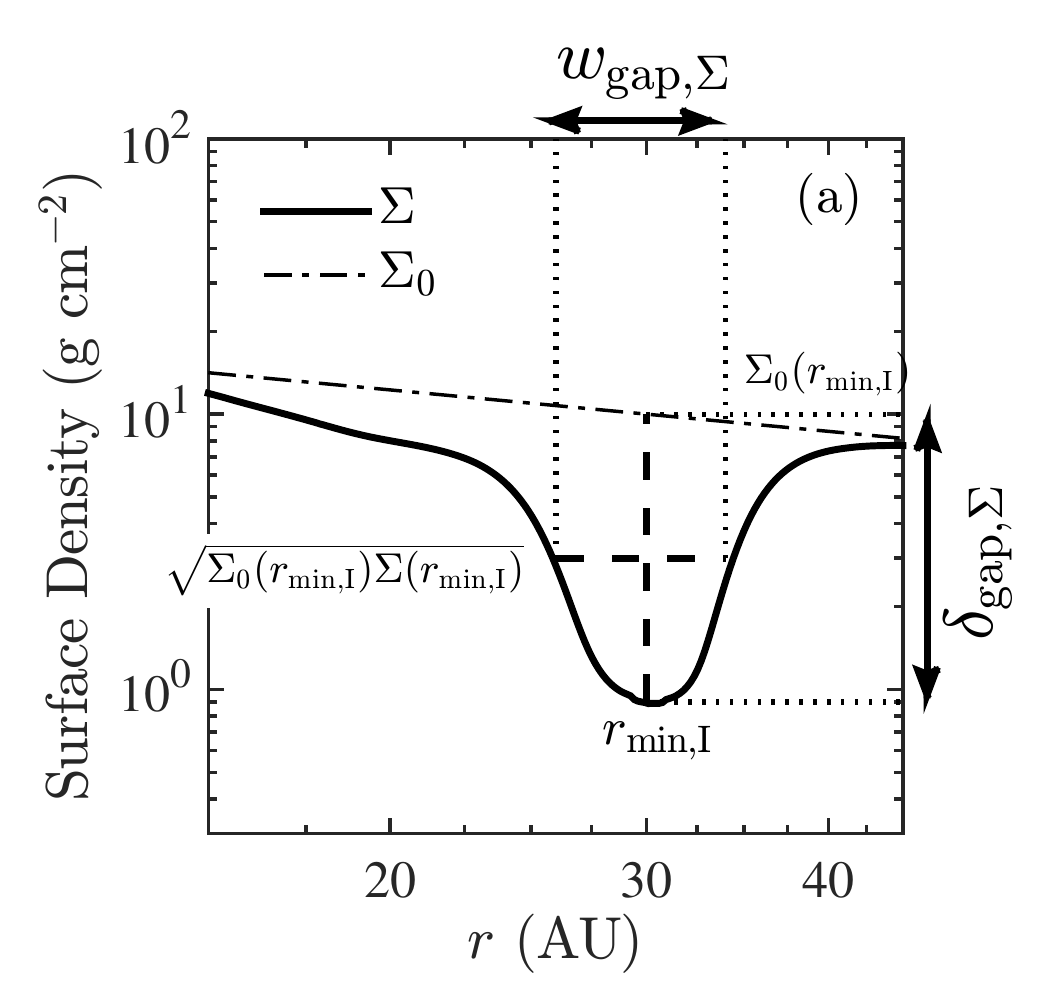}
\includegraphics[trim=0 0 0 0, clip,width=0.49\textwidth,angle=0]{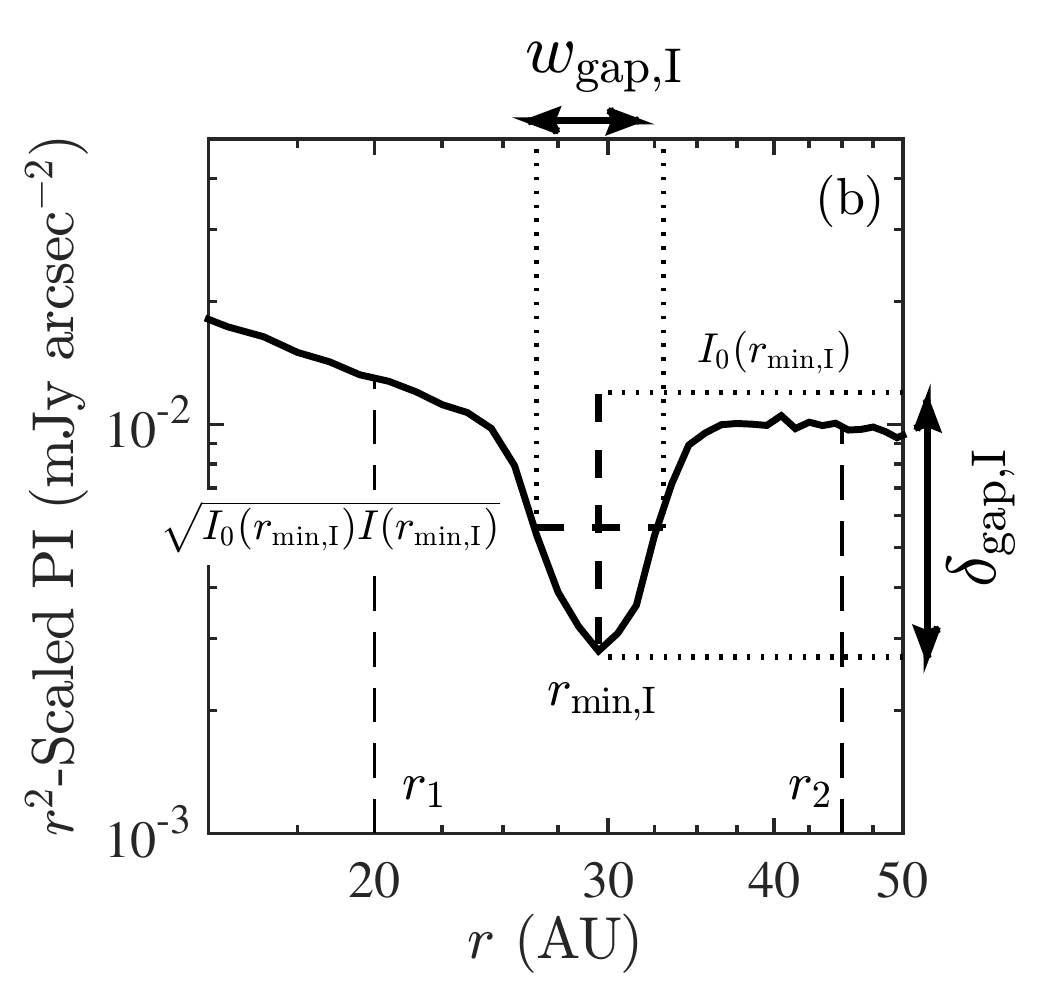}
\end{center}
\figcaption{Definitions of $\widthsigma$ and $\depthsigma$ in surface density (a), and $\widthimage$ and $\depthimage$ in scattered light image (b). The gap depth (the vertical dashed segment) is defined as the ratio between an ``undepleted background'' and the minimum value of the quantity inside the gap at radius $\rmin$ ($\rminsigma$ or $\rminimage$; $\approx\rp=$30 AU): $\depthsigma=\Sigma_0(\rminsigma)/\Sigma(\rminsigma)$ and $\depthimage=I_0(\rminimage)/I(\rminimage)$ ($I$ is the $r^2$-scaled image surface brightness). $\Sigma_0(\rminsigma)$ is the initial $\Sigma_0$ (dash-dotted line) at $\rminsigma$; $I_0(\rminimage)$ is taken to be the geometric mean of $I$ at $r_1=(2/3)\rminimage$ and $r_2=(3/2)\rminimage$ (marked in (b)). $r_1$ and $r_2$ are far away from the gap, and for narrow gaps, planet-induced perturbation should be small at these distances. The gap width (horizontal dashed segment) is defined as the distance between the inner and outer edges of the gap (two vertical dotted lines), at which radius the quantities reaches the geometric mean of the background and the gap minimum (i.e., $\sqrt{\Sigma_0(\rminsigma)\Sigma(\rminsigma)}$ and $\sqrt{I_0(\rminimage)I(\rminimage)}$).
\label{fig:definition}}
\end{figure}

\begin{figure}
\begin{center}
\includegraphics[trim=0 0 0 0, clip,width=0.49\textwidth,angle=0]{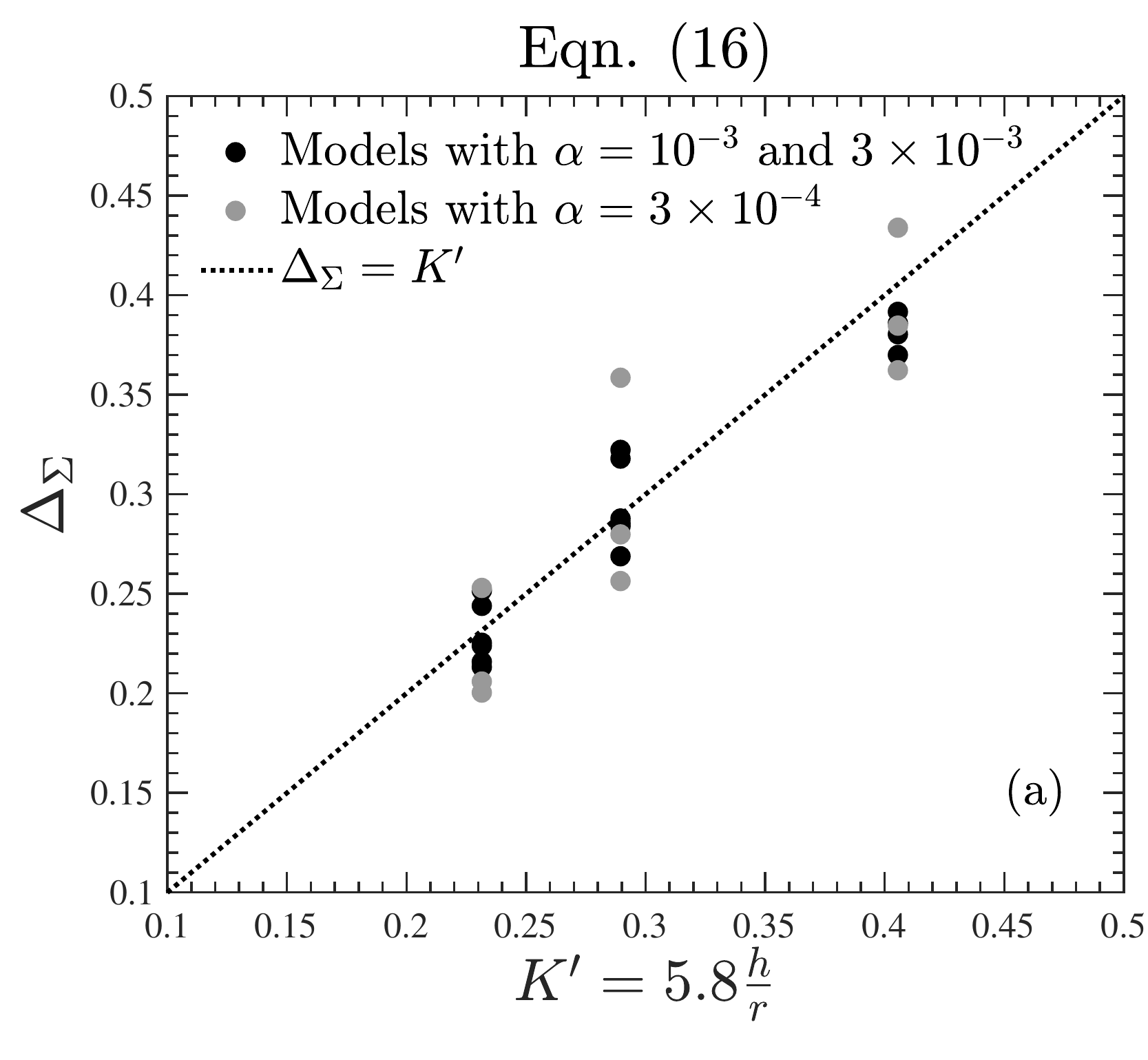}
\includegraphics[trim=0 0 0 0, clip,width=0.49\textwidth,angle=0]{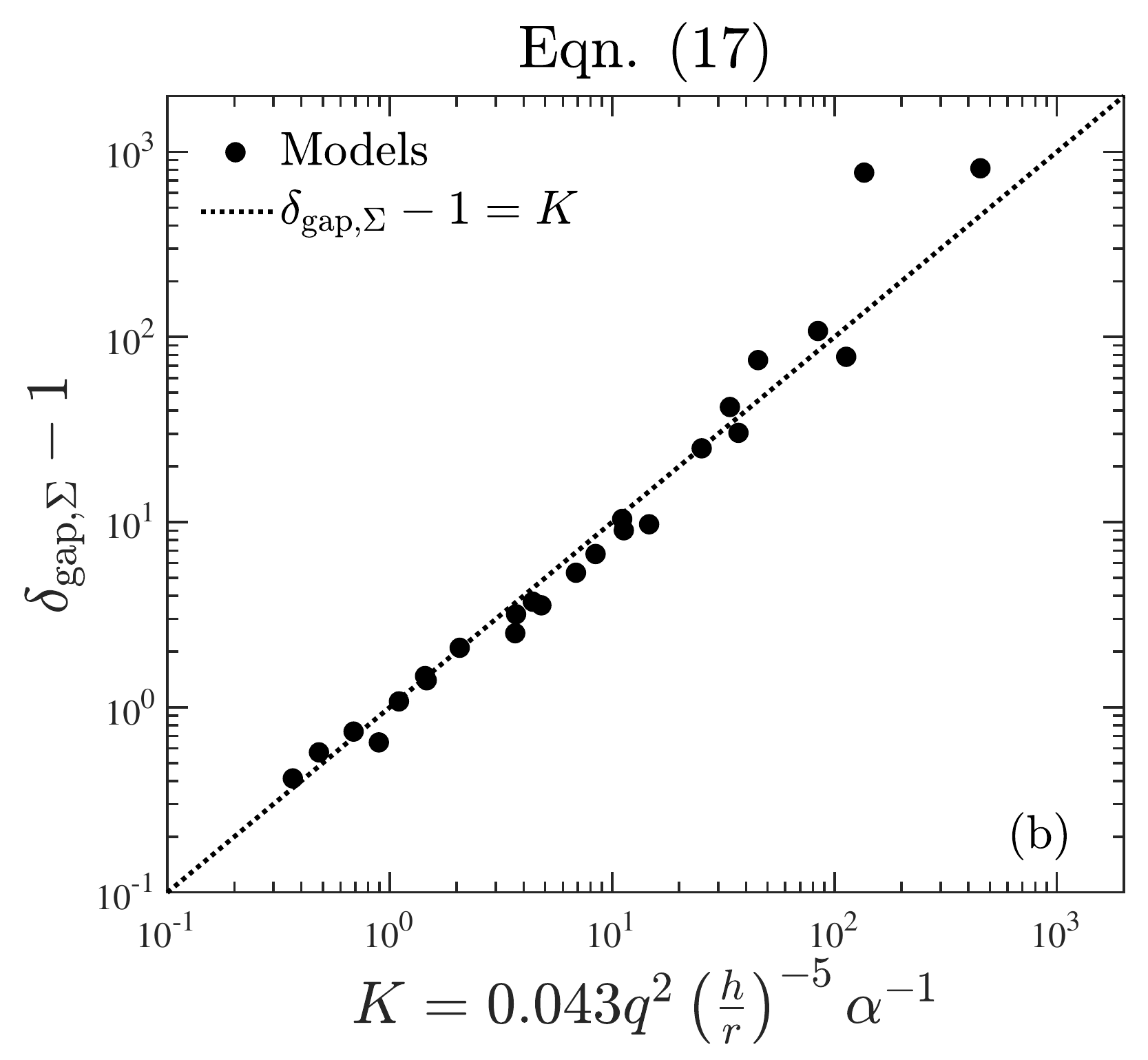}
\end{center}
\figcaption{The correlations between gap width (a) and depth (b) in surface density and the three planet and disk parameters --- $q=\mplanet/M_\star$, $h/r$, and $\alpha$ --- as described in Equations~\ref{eq:width-sigma-h} and \ref{eq:depth-sigma}. See Section~\ref{sec:sigma} for details.
\label{fig:width_depth_sigma}}
\end{figure}

\begin{figure}
\begin{center}
\includegraphics[trim=0 0 0 0, clip,width=0.49\textwidth,angle=0]{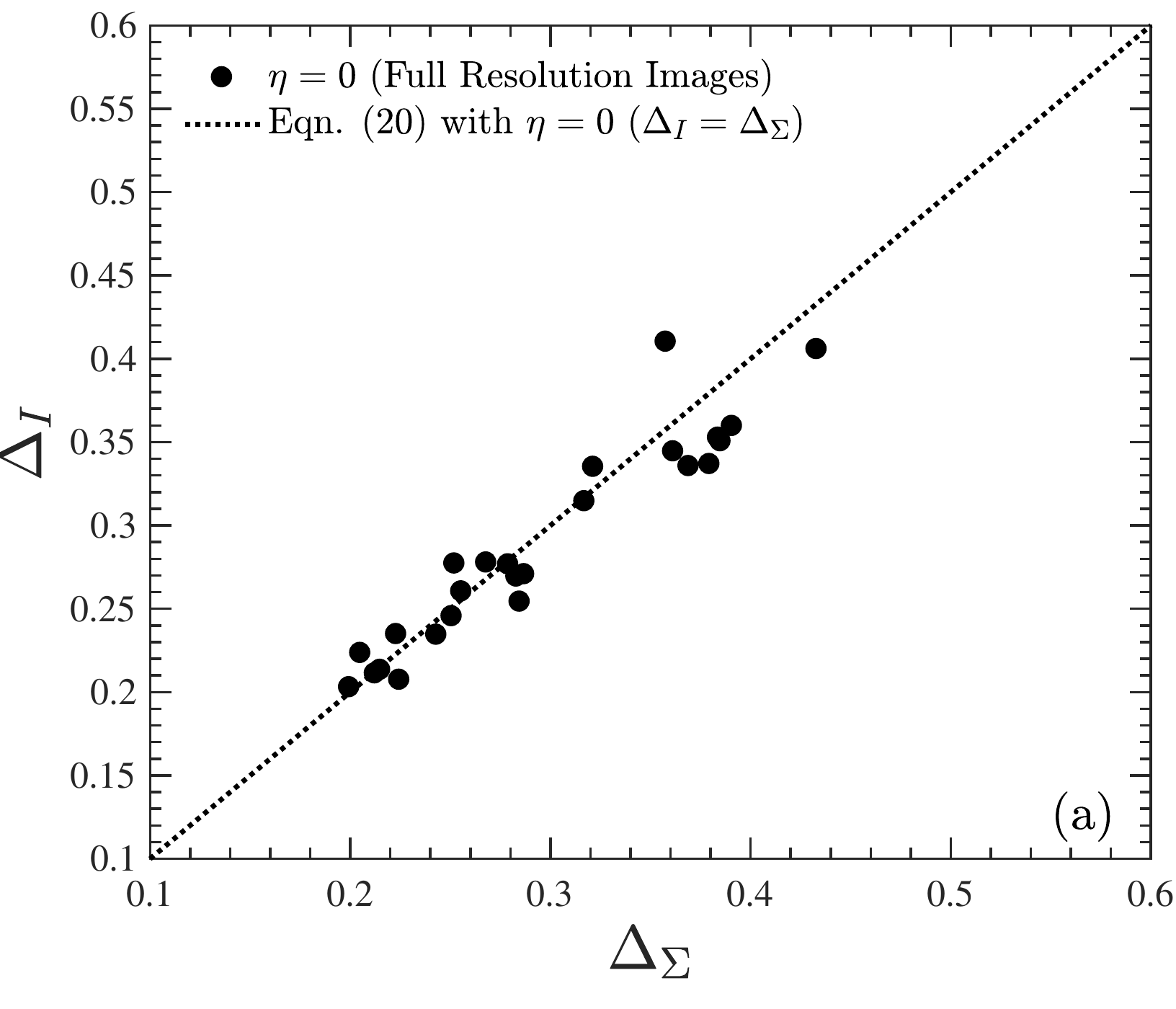}
\includegraphics[trim=0 0 0 0, clip,width=0.49\textwidth,angle=0]{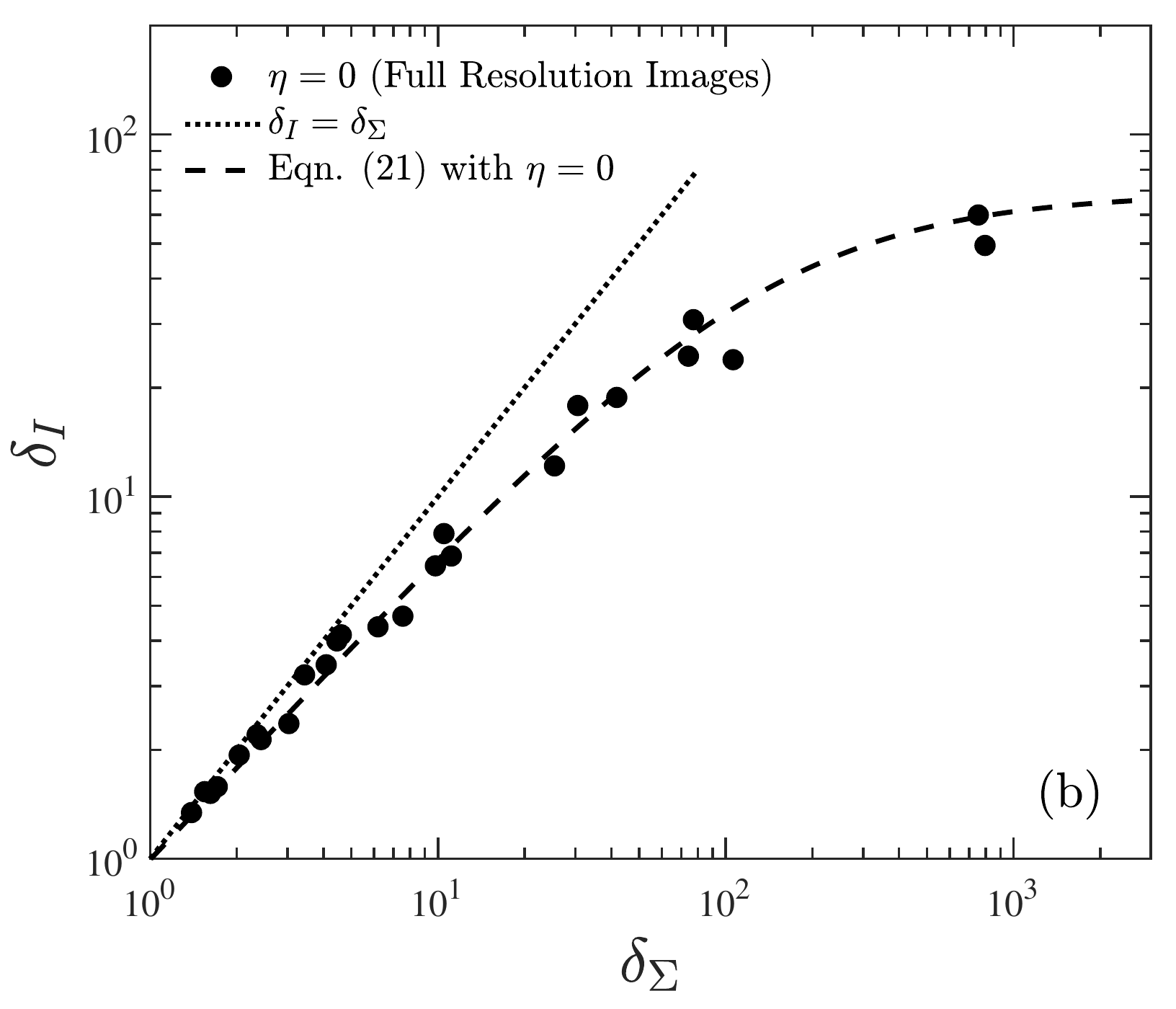}
\end{center}
\figcaption{Gap width in scattered light {\it vs} gap width in surface density, and gap depth in scattered light {\it vs} gap depth in surface density (b). See Section~\ref{sec:faceon} for details.
\label{fig:width_depth}}
\end{figure}

\begin{figure}
\begin{center}
\includegraphics[trim=0 0 0 0, clip,width=0.49\textwidth,angle=0]{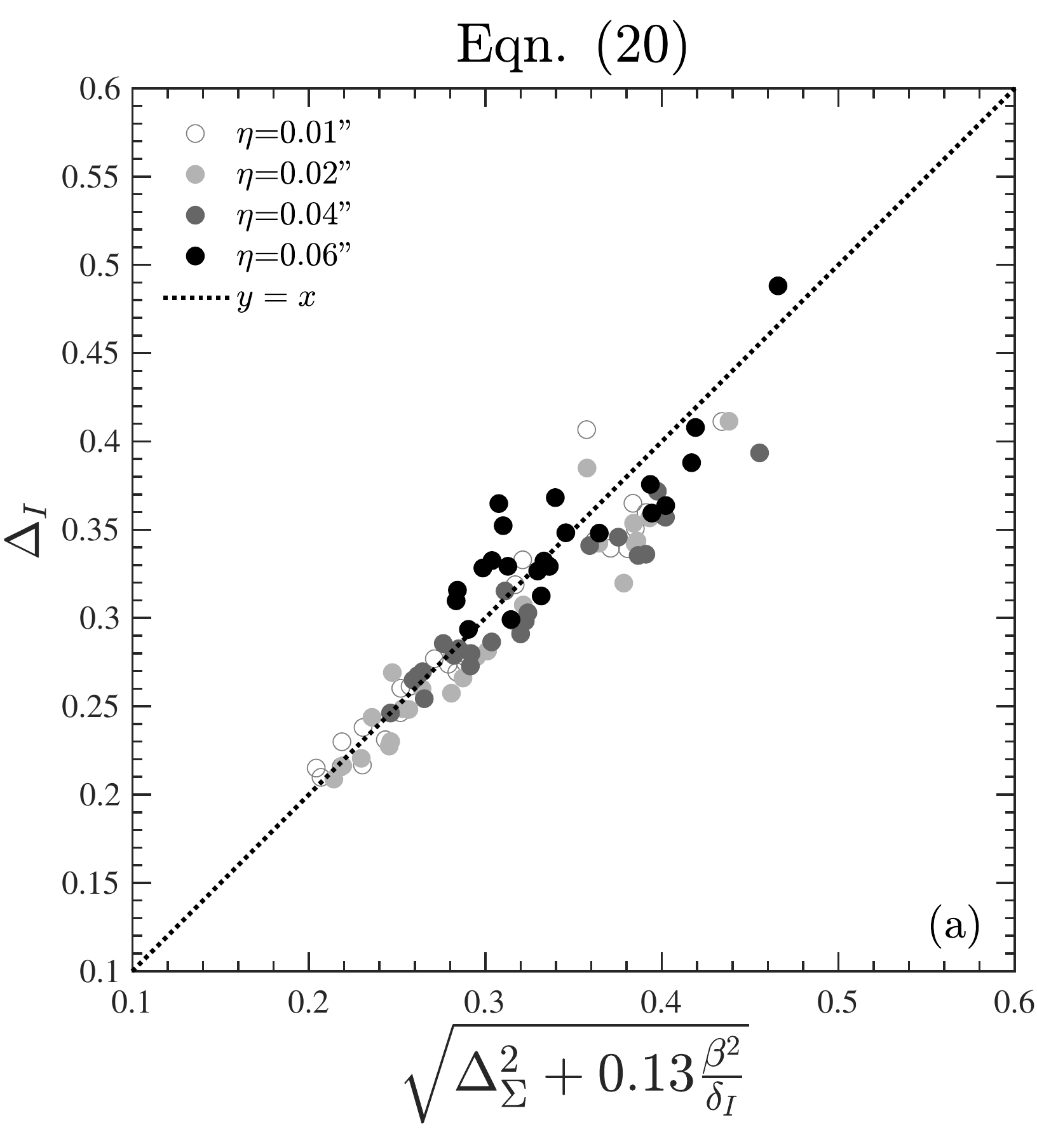}
\includegraphics[trim=0 0 0 0, clip,width=0.49\textwidth,angle=0]{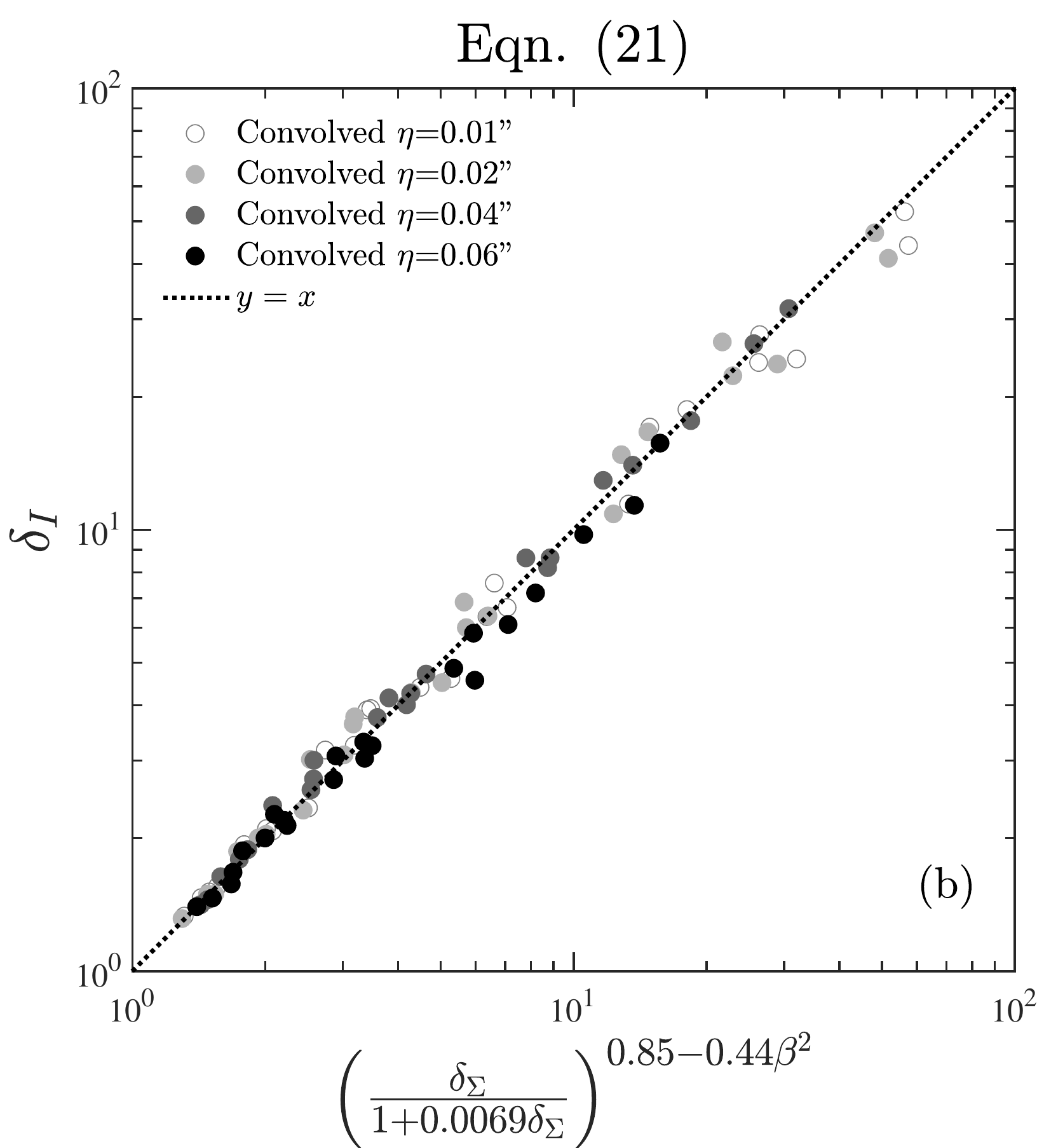}
\end{center}
\figcaption{The effect of PSF smearing in gap width and depth, showing the correlations described by Equations~\ref{eq:width-image-sigma} and \ref{eq:depth-image-sigma}. See Section~\ref{sec:faceon} for details.
\label{fig:width_depth_raw_conv}}
\end{figure}

\begin{figure}
\begin{center}
\includegraphics[trim=0 0 0 0, clip,width=0.66\textwidth,angle=0]{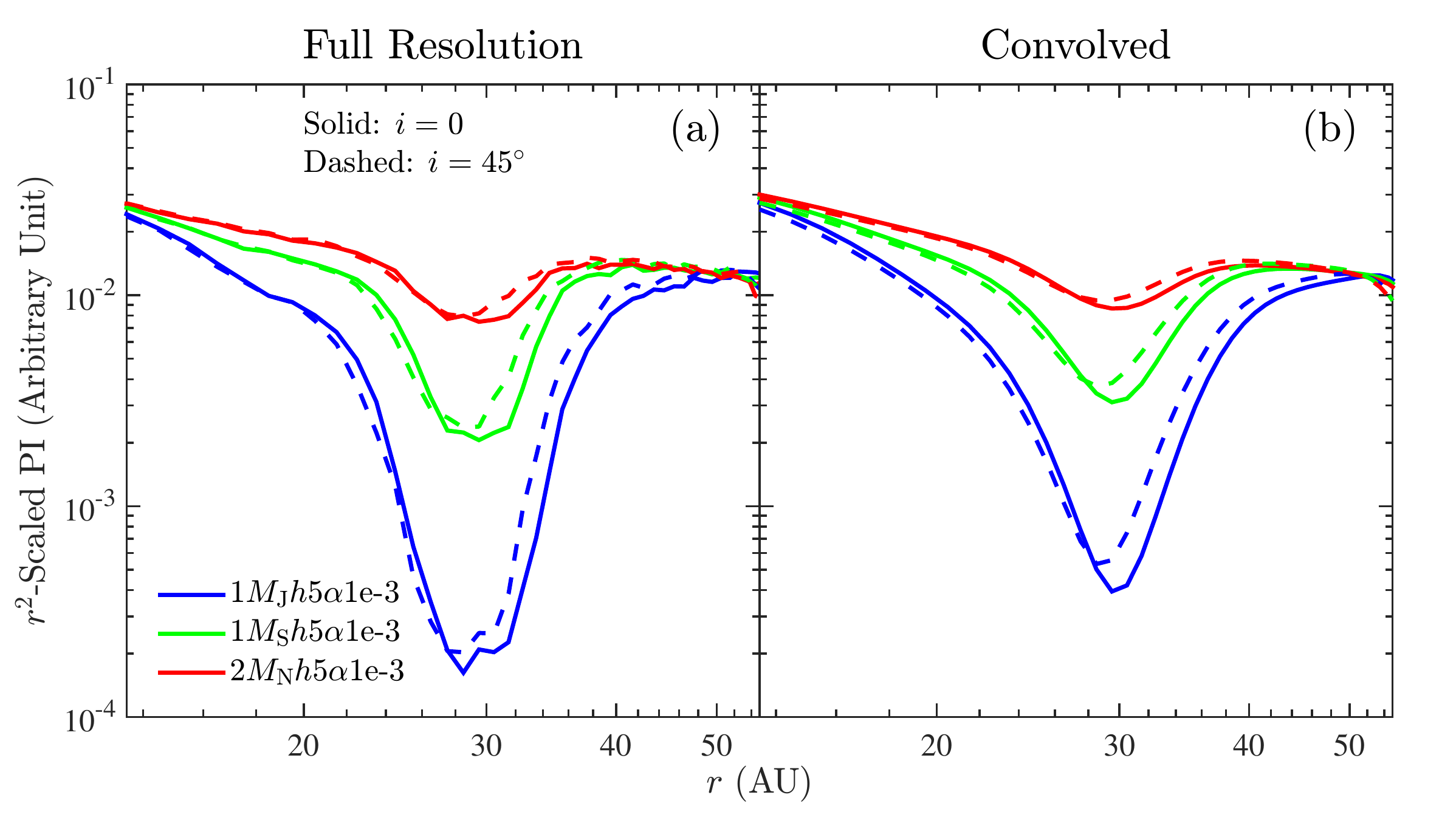}
\includegraphics[trim=0 0 0 0, clip,width=0.33\textwidth,angle=0]{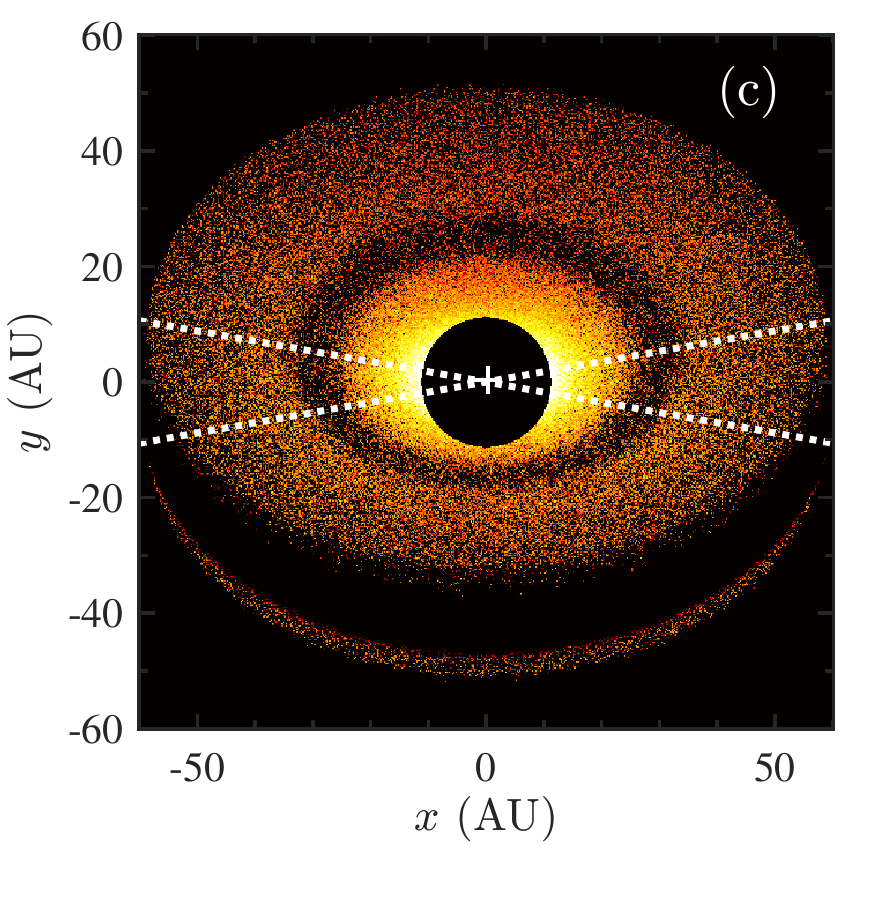}
\end{center}
\figcaption{Radial profiles of three representative models with $h/r=0.05$, $\alpha=10^{-3}$, and $\mplanet=1\mj$, $1\ms$, and $2\mn$ (red, green, and blue curves, respectively). Solid curves are azimuthally averaged radial profiles for face-on images; dashed curves are radial profiles along the major axis at $i=45^\circ$, averaged over a wedge with an opening angle of 20$^\circ$, as indicated in panel (c). Panels (a) and (b) are for full resolution and convolved images, respectively. All $i=45^\circ$ radial profiles have been scaled by the same constant factor to meet the solid curves at the same point at 12 and 55 AU. Radial profiles at $45^\circ$ are very similar to radial profiles at face-on. See Section~\ref{sec:inclinations} for details.
\label{fig:image_rp_h5a1e3_45}}
\end{figure}

\begin{figure}
\begin{center}
\includegraphics[trim=0 0 0 0, clip,width=\textwidth,angle=0]{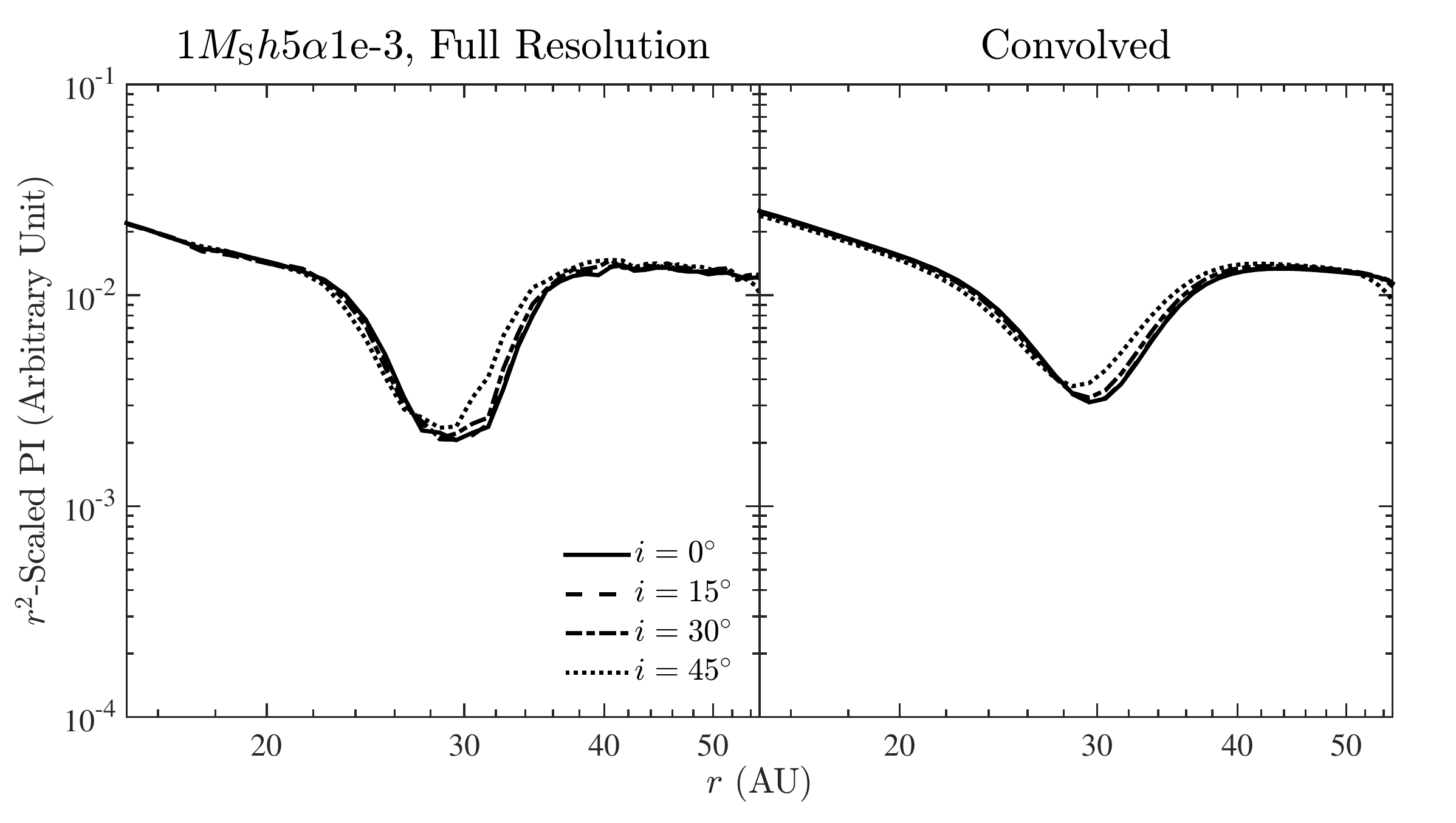}
\end{center}
\figcaption{Radial profiles of the 1$M_{\rm S}h$5$\alpha$1e-3 model in full resolution {\bf (left)} and with $\eta=0\arcsec.04$ {\bf (right)} at 4 different inclinations. For $i=0$ radial profiles are azimuthally averaged, while at finite $i$ radial profiles are averaged over a wedge with an opening angle of $20^\circ$ (Figure~\ref{fig:image_rp_h5a1e3_45}c). The curves with $i\ne0$ have been scaled so that they meet the $i=0$ profile at 12 and 55 AU. The gap depth and width at $i\leq45^\circ$ depend on inclinations only weakly. See Section~\ref{sec:inclinations} for details.
\label{fig:image_rp_1msh5a1e3_i}}
\end{figure}

\begin{figure}
\begin{center}
\includegraphics[trim=0 0 0 0, clip,width=0.49\textwidth,angle=0]{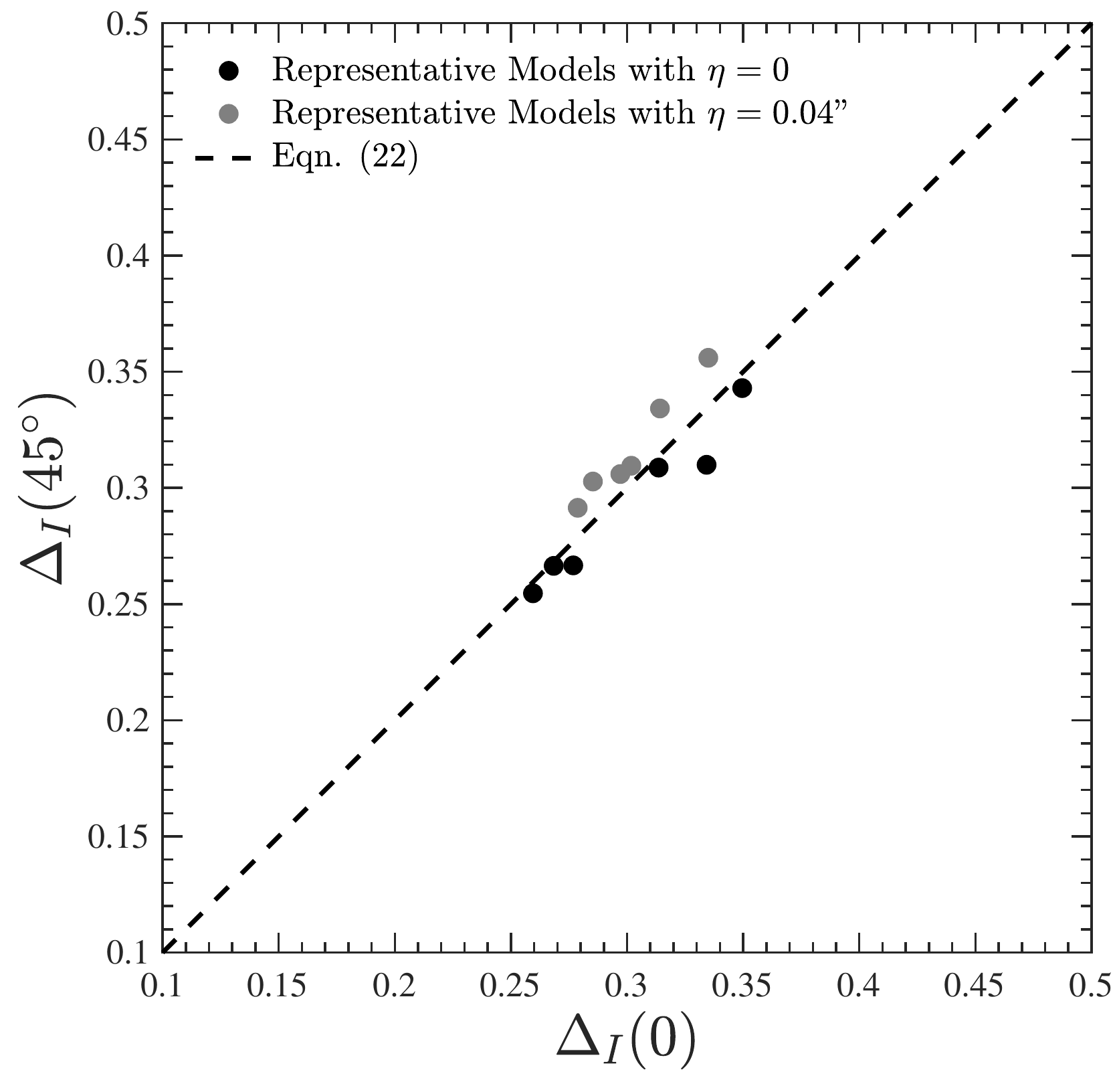}
\includegraphics[trim=0 0 0 0, clip,width=0.49\textwidth,angle=0]{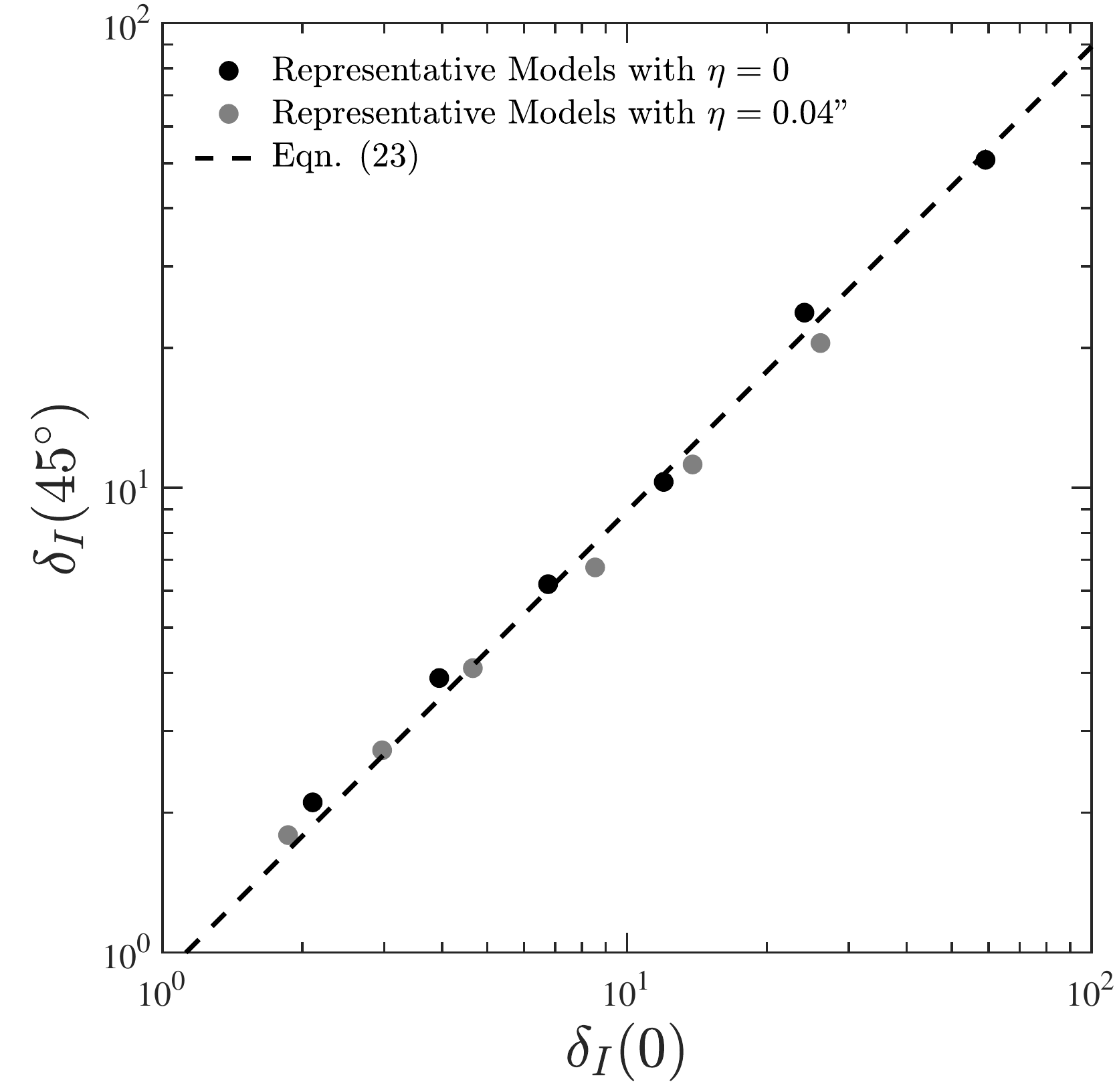}
\end{center}
\figcaption{Gap width and depth in inclined disks. Panel (a) shows the gap width at $i=45^\circ$ {\it vs} $i=0$ for 6 representative models for both full resolution (black points) and convolved images (gray points). The models are 1$\mj h$5$\alpha$3e-3,  1$\mj h$5$\alpha$1e-3, 1$\mj h$7$\alpha$1e-3, 1$\ms h$5$\alpha$1e-3, 2$\mn h$5$\alpha$1e-3, and 2$\mn h$5$\alpha$3e-4 Panel (b) shows the gap depth at $i=45^\circ$ {\it vs} $i=0$ for these models. Correlations~\ref{eq:width-i} and \ref{eq:depth-i} are overplotted. 
\label{fig:depth_depth_45}}
\end{figure}

\begin{figure}
\begin{center}
\includegraphics[trim=0 0 0 0, clip,width=\textwidth,angle=0]{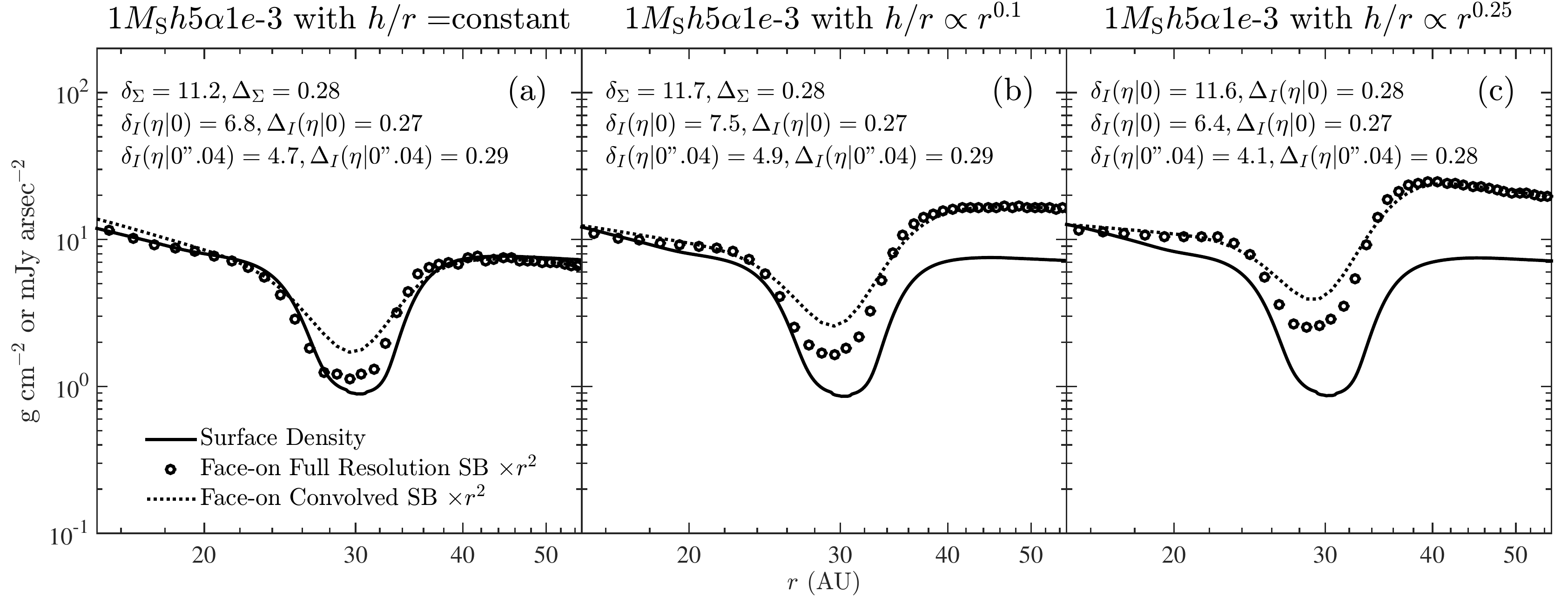}
\end{center}
\figcaption{Radial profiles of the surface density (solid) and $H$-band polarized intensity surface brightness (circles: full resolution; dotted line: convolved; both scaled by $r^2$) of three models with $h/r=0.05$ at $\rp=30$ AU, $\alpha=10^{-3}$, $\mplanet=1\ms$, and different radius dependence of $h/r$: {\bf (a)} $h/r$ is a constant (flat disk; our fiducial setting), {\bf (b)} $h/r\propto r^{0.1}$, {\bf (c)} $h/r\propto r^{0.25}$. The disk in (b) and (c) is flared. Both the full resolution and convolved surface brightness curves in each panel have been scaled by the same constant factor so that they roughly meet the surface density curves at $r=15$ AU. As the disk becomes more flared, the outer disk becomes brighter, as expected. However, the width and depth of the gap in both the surface density and scattered light (measurements printed in each panel) stay roughly the same, as the gap is approximately a local structure, thus not significantly affected by the global flareness of the disk. See Section~\ref{sec:flareness} for details.
\label{fig:sigma_image_diffhrpower_rp}}
\end{figure}

\begin{figure}
\begin{center}
\includegraphics[trim=0 0 0 0, clip,width=\textwidth,angle=0]{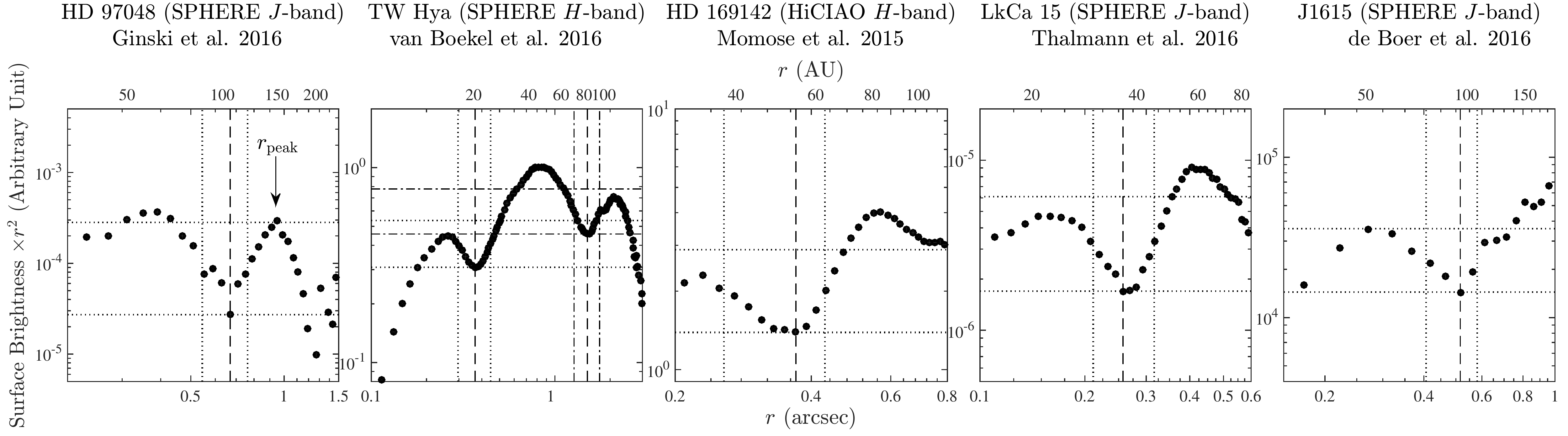}
\end{center}
\figcaption{Radial profiles of $r^2$-scaled polarized intensity (arbitrary unit) of HD 97048 (along the major axis on the north), TW Hya (azimuthally averaged), HD 169142 (azimuthally averaged after deprojecting the disk as in \citealt{momose15}), LkCa~15 (major axis; averaged over the two sides), and J1615 (major axis; averaged over the two sides). In each panel, the vertical dashed-line indicates $\rminimage$; the two horizontal dotted lines indicate $I_0$ and $I_{\rm min}$; and the two vertical dotted lines indicate $r_{\rm out}$ and $r_{\rm in}$ (in TW Hya, we use two horizontal dash-dotted lines and two vertical dash-dotted lines to indicate $I_0$, $I_{\rm min}$, $r_{\rm out}$, and $r_{\rm in}$ for the outer gap). In HD 97048, the outer disk point ``$r_2$'' in the gap definition (Figure~\ref{fig:definition}) is significantly outside the radius of the peak in the outer disk ($r_{\rm peak}=0\arcsec.95$; indicated by the arrow); we thus fix $r_2=r_{\rm peak}$ as the ``outside the gap'' point. $r_{\rm out}$, $r_{\rm in}$, $I_0$, and $I_{\rm min}$ are listed in Table~\ref{tab:applications}. See Section~\ref{sec:applications} for details.
\label{fig:real_disks}}
\end{figure}

\end{document}